# Phase Behavior of Thermo-Responsive Nanoplatelets

Imane Boucenna,[*] Florent Carn[*] and Ahmed Mourchid[*]

Matière et Systèmes Complexes (MSC), UMR 7057 CNRS and Université Paris Cité, 10 rue Alice Domon et Léonie Duquet, 75205 Paris Cedex 13, France

**Abstract.** Nanoplatelets open up a wide range of possibilities for building materials with novel properties linked to their shape anisotropy. A challenge consists of controlling dynamically the order of positioning and orientation in three dimensions by assembly to exploit the collective properties at the macroscale. While most studies to date have focused on hard platelets that cannot be stimulated by an external trigger, in the present work we tackle the case of core/shell platelets, composed of a mineral core (laponite) coated with a soft polymeric shell ($PEO_{100}$-$PPO_{65}$-$PEO_{100}$ copolymer) whose swelling can be triggered by temperature variation. We identified unambiguously the signature associated with the different phases obtained as a function temperature and concentration by combining local (X-ray scattering, electron microscopy) and global (rheology, optical birefringence) methods of analysis. In this way, we obtained two main results. The first shows that the deposition of a soft layer onto laponite surface enables a phase transition from isotropic liquid towards liquid suspensions of random stacks which is not observed for bare laponite suspensions in the studied weight concentration range ($\phi \leq 14$ wt.%). The second result shows that an increase of nanoplatelet effective volume fraction triggered by temperature (swelling of the polymer shell) induces a phase transition from liquid suspensions of random stacks towards birefringent gels of nematic stacks. Both results agree with numerically predicted phase sequences expected by variation in particle density under similar charge screening conditions, taking into account the contribution of the copolymer layer to the particle volume fraction. We believe that these results pave the way for the control of nanoplatelet self-assembly by external action.

# Introduction

Two-dimensional nanoparticles (i.e. nanoplatelets) represent a vast class of nanoparticles, which can be of natural or synthetic origin, and are used in a wide range of applications. This range of applications has grown in recent years with the possibility of self-assembling these nanoplatelets to form advanced macroscopic materials with new properties linked to their shape and collective properties. This has been demonstrated in single-photon emitting devices [1], electrode materials [2], solar concentrators [3] or photo-catalytic devices [4]. These developments have led to renewed interest in studying the phase behavior of nanoplatelets, with the aim of better understanding and controlling the emergence of a three-dimensional order of positioning and orientation [5-10]. For example, recent studies explored the possibility of obtaining new structures from geometrically frustrated assemblies by playing on the effect of particle confinement or particle shape. In the same time, the phase behavior of charged rigid platelets in solution is still a subject of attention both from an experimental and a theoretical point of view. With regard to the first aspect, one difficulty lies in the experimental difficulty of unambiguously identifying certain liquid crystal phases (i.e. cubatic phase, lamellar vs. columnar nematic phases) [6]. From a theoretical point of view, recent developments concern the prediction of phase diagrams by Monte Carlo simulation, taking into account the interaction between shape anisotropy and the electrostatic interactions of charged platelets [11]. This work has predicted a phase diagram even richer than that previously expected on the basis of simple excluded volume interactions, which not only includes crystalline or liquid crystalline phases, but also a new structure, called inter-growth texture, which has not yet been observed experimentally.

In this context, we investigated the possibility of controlling nanoplatelet assembly by adsorbing a flexible layer of thermosensitive non-ionic copolymer on their surface. Our initial objective is to show that a phase transition can be induced by varying the volume fraction of the coated platelets using an external trigger, in this case a temperature variation affecting the copolymer conformation. Although the influence of polymer adsorption on platelets phase behavior has been studied [12], the influence of a thermosensitive polymer layer has not yet been considered.

Our system is an aqueous suspension of core/shell nanoplatelets composed of laponite particles covered by a soft shell of non-ionic copolymer. Laponite is a synthetic smectite clay having a thickness of about 1 nm and an average diameter $D \approx 28$ nm with a polydispersity of 30 % [13, 14]. In our case, the platelets interact essentially through excluded-volume interactions, as electrostatic interactions between platelets are screened due to the presence of a counter-ion, $Na_4P_2O_7$, which yield a Debye length ($\kappa^{-1}$) comparable to the platelet thickness. Moreover, the presence of these counter-ions facilitates platelet dispersion by screening the positive edge charges of the laponite. It results in a drastic reduction of the suspension viscosity as compared to suspension of laponite without $Na_4P_2O_7$ counter-ions by avoiding the formation of the house of cards structure [15, 16]. The non-ionic copolymer, is a triblock copolymer poly(ethylene oxide)$_{100}$/poly(propylene oxide)$_{65}$/poly(ethylene oxide)$_{100}$, noted PEO-PPO-PEO (e.g. pluronic F127). The thermosensitive character of this macromolecule is due to thermal changes in the degree of hydrophilicity of both PEO and PPO blocks. This results in two successive processes when the temperature is increased from ~ 5 °C: (i) self-association of the PPO blocks, above a critical temperature (T*), while PEO blocks remain exposed to water; (ii) swelling of the PEO brush when the temperature increases further. This behavior has been first observed in binary systems of pluronic molecules and water and then identified in ternary systems composed of laponite coated with these molecules in water [17-19]. In the first case, the molecules self-assemble into micelles above T*, whereas in the second case, one observes the formation of core-shell particle. In this study, we explored the phase behavior of these core-shell nanoplatelets as a function of laponite concentration and of temperature by means of small angle X-ray scattering (SAXS), cryogenic transmission electronic microscopy (cryo-TEM), birefringence observation and rheology. Using this multi-technique approach, we have identified the specific signature associated with each of the phases observed in the different concentration and temperature ranges.

Overall, this work has enabled us to obtain the following two main results. The first is to show that the change in aspect ratio obtained by the deposition of a soft polymer layer on the laponite surface leads to a phase transition from isotropically dispersed nanoplatelet suspensions towards liquid suspensions of random stacks of nanoplatelets. This result is quite similar to previous numerical simulations and experimental phase behavior mapping data of hard nanoplatelets [20-24]. The second result is to show that a variation in core-shell nanoplatelets effective volume fraction, induced by an increase in temperature, leads to a phase transition from liquid suspensions of random stacks towards birefringent gels composed of nematic stacks of nanoplatelets. Both experimental results are in agreement with the most recent numerical predictions developed for hard platelets interacting with a screened Coulomb potential in the limit where the screening parameter is rather high (i.e. where the product $\kappa D \geq 14$) [11, 25].

## Experimental Section

**Materials.** We used a block copolymer poly(ethylene oxide)$_{100}$/poly(propylene oxide)$_{65}$/poly(ethylene oxide)$_{100}$, with a nominal weight averaged molar mass of 12600 g/mol (Sigma-Aldrich). The synthetic clay used was laponite RDS, obtained as a gift from Rockwood Additives (now BYK Additives). Laponite RDS differs from laponite RD, which has been often used in prior studies [16], by the presence a so called peptizing agent (e.g. tetrasodium pyrophosphate, [Na$_4$P$_2$O$_7$] ≈ 8.2 g per 100 g of dried Laponite according to the supplier) that allow the formation of liquid suspensions at high laponite concentration. This effect results from the dissociation of the peptizing agent in water (4Na$^+$, P$_2$O$_7^{4-}$) and the adsorption of pyrophosphate anions (P$_2$O$_7^{4-}$) on the positive charges located on the edge of the laponite. As a result, the nanoplatelets behave like negatively charged particles in a relatively salty solution.

**Sample preparation.** The samples were prepared by mixing the solid powder with aqueous solutions at fixed pH = 10 with NaOH. The ionic strength, $I = \frac{1}{2}\sum_i c_i Z_i^2$, and thus the Debye length, $\kappa^{-1}$, of the suspensions could be estimated by only taking into account the contribution of the excess Na$_4$P$_2$O$_7$ as shown in the section 7 of the supporting information. In our experiments, we estimate (see table S9) that the variation in laponite mass fraction (ϕ) from 1% to 14% is associated with a variation in Debye length from approximately 2.3 to 0.6 nm, assuming that only pyrophosphate counter ions that don't interact with Laponite platelets contribute to the ionic strength [26]. As a consequence, electrostatic interactions are screened for the concentrations investigated in this study with screening parameter, κD, ranging from 14 (ϕ = 1%) to 53 (ϕ = 14%). The samples were prepared at a fixed ratio of laponite/copolymer (i.e. 0.81 g of copolymer per g of dry laponite powder) for which the copolymer adsorbs onto the laponite to produce hairy platelets without copolymer micelles in excess as already shown in the past [18, 27, 28]. According to small angle neutron scattering (SANS) measurements (figure S1 of supporting information (SI)), the adsorbed layer of copolymer has a thickness △e = 32 ± 1.0 Å, which is assumed to be uniform on both the edge and sides of all the nanoplatelets thanks to nonspecific dipolar interactions and to the high chemical homogeneity of Laponite synthetic platelets. These values are consistent with previous SANS measurements [18, 28]. This thickness is mainly associated with the PEO blocks exposed to the solvent, while the PPO blocks are adsorbed directly onto the nanoparticles. Overall, these characteristics lead to an effective thickness and diameter for the coated laponite of 73 and 344 Å respectively, giving a modified aspect ratio l/D ~ 0.21 (i.e. 6 times larger than for bare laponite). In addition, the temperature dependence of the PEO chains conformation in water allows the volume fraction of the coated platelets to vary with a minimal influence on the aspect ratio [29]. The concentration of each species, i.e., nanoplatelet, copolymer, total pyrophosphate (present in the raw material) and calculated free pyrophosphate in each solution are given in table S9 of the SI (section 7), as well as the variation of ionic strength and Debye lengths with the Laponite concentration.

**Small angle X-ray scattering (SAXS).** SAXS measurements were performed on the Swing beamline at synchrotron Soleil (France) at a photon energy of 12 keV and a wavelength of 1.03 Å. The measurements were acquired during different runs at the facility. During each of them, the scattering patterns on each sample, prepared the previous week, and on water in capillary tubes of 1.5 mm, as well as on an empty tube, were measured at a fixed duration time and radially averaged to obtain the scattering intensity as a function of the norm of the wavevector, I(q). Then the scattering background considered to be water in the capillary tube was subtracted from the data. The absolute scale (cm$^{-1}$) was obtained by normalization with respect to the average (flat) scattering signal of ($I_{\text{water in the capillary}} - I_{\text{capillary}}$) around q = 0.1 Å$^{-1}$ which is known to be 0.0163 cm$^{-1}$ (water differential scattering cross section). There is a large difference of the contrast factor between nanoparticles and coating polymer. Laponite, copolymer and water scattering length densities, $\rho_{lap}$, $\rho_{pol}$ and $\rho_{\text{water}}$, are 2.280×10$^{-5}$ Å$^{-2}$, 0.975×10$^{-5}$ Å$^{-2}$ and 0.945×10$^{-5}$ Å$^{-2}$ respectively. Thus the copolymer contribution to X-rays scattering is at least 2 orders of magnitude weaker than the signal of the nanoparticles: $(\rho_{pol} - \rho_{\text{water}})^2$ = 0.0009×10$^{-10}$ Å$^{-4}$ as against $(\rho_{lap} - \rho_{\text{water}})^2$ = 1.78×10$^{-10}$ Å$^{-4}$.

**Cryogenic transmission electron microscopy (cryo-TEM).** Sample preparation and observation by cryo-TEM were realized according to the following procedure. First, a (low viscosity) sample droplet was placed on a carbon membrane grid and the excess of sample was quickly removed by a filter paper, leaving a thin film on the grid, then it was immediately quench-frozen in liquid ethane and cooled with liquid nitrogen. After verification that sample thickness was adequate, its observation was achieved with a LaB6 JEOL JEM 2100 electron microscope. The observations were done at the platform of the Institute of Mineralogy, Physics of Materials and Cosmochemistry (IMPMC, Paris).

**Rheology.** The evolution of elastic and viscous moduli, G' and G" respectively, was measured as a function of temperature at a fixed frequency of 1 Hz, an applied strain of 0.2 %, and a heating rate of 1 °C/min. The measurements were carried out on a Physica RheoCompass rheometer by using a cone and plate geometry having a diameter of 5 cm and cone angle of 1° fitted with a solvent trap to minimize water evaporation.

**Birefringence.** Birefringence of the samples was observed between cross-polarizers on samples previously heated to the measuring temperature.

## Results and Discussion

### Phase behavior of bare and coated laponite at T = 22 °C.

In this first part, we studied the structure of bare and copolymer coated nanoplatelets suspensions at room temperature as a function of concentration ($\phi$ = 1 wt. % → 14 wt. %). Figures 1a and S2 of supporting information (SI) show the curves obtained for bare laponite. At the lowest concentration, the experimental curves correspond to the theoretical form factor of polydisperse discs (thickness = 0.92 nm and diameter, D = 28 nm with a log-normal polydispersity = 30 %) on the whole q range. When the laponite concentration is increased, the scattered intensity, I(q), decreases progressively with respect to the disc form factor in the low-q domain where $q < 2\pi/D$. This evolution reflects the presence of excluded volume interactions as previously shown in reference [30]. However, we did not observe either a correlation peak or a permanent birefringence in the studied concentration range for bare laponite. Moreover, all samples are in a low-viscosity liquid state throughout the concentration range during the course of the measurements. All these observations agree with previous studies of bare laponite [13, 14, 31]. Let's now turn to the case of coated laponite suspensions, considering the SAXS curves shown in Figures 1b and S1. It appears that the scattering curves of coated laponite are similar to those obtained for bare laponite at low concentrations. However, they become progressively different as the particle concentration increases with the progressive emergence of a structural peak located at $q^* \sim 0.03$ Å$^{-1}$ (i.e. d = $2\pi/q^*$ ~ 20 nm) when $\phi \geq 2$ wt. %. The amplitude of the peak increases with particle concentration, and a second order peak gradually emerges at $q^{**} \sim 0.06$ Å$^{-1}$. To better identify these peaks, we have plotted in figures 1c and 1d the scattering curves of bare laponite and of coated laponite respectively, in Kratky representation (i.e. Iq² *vs.* q). In this representation, Iq² is proportional to the structure factor since the form factor of nanoplatelets decays as $q^{-2}$ when $q \geq 2\pi/D \sim 0.02$ Å$^{-1}$ [12]. Thus, one could observe that the position of the peak does not show significant shift when $\phi$ increases. These measurements were completed by observations with a cryogenic transmission electronic microscope (cryo-TEM, Figure 2) which enabled to identify the appearance of laponite stacking when concentration exceed 2 wt. % (Figures 2c-d). Quantitative measurement of characteristic lengths for a 3D system from 2D images is tricky, and even more when considering platelets as thin as laponite (thickness ~ 1 nm). Nevertheless, it appears that the inter-particle distances observed are comparable with the characteristic distance derived from the SAXS structure peak (i.e. $q \sim 0.03$ Å$^{-1}$ → $d \sim 20$ nm). This distance is of the order of magnitude expected for two layers of 100-monomer PEO in a good solvent (i.e. $2 \times 0.35 \times 100^{0.6} \sim 12$ nm) bearing in mind that the PPO adsorbed on the surface of the platelets adds a thickness [32].

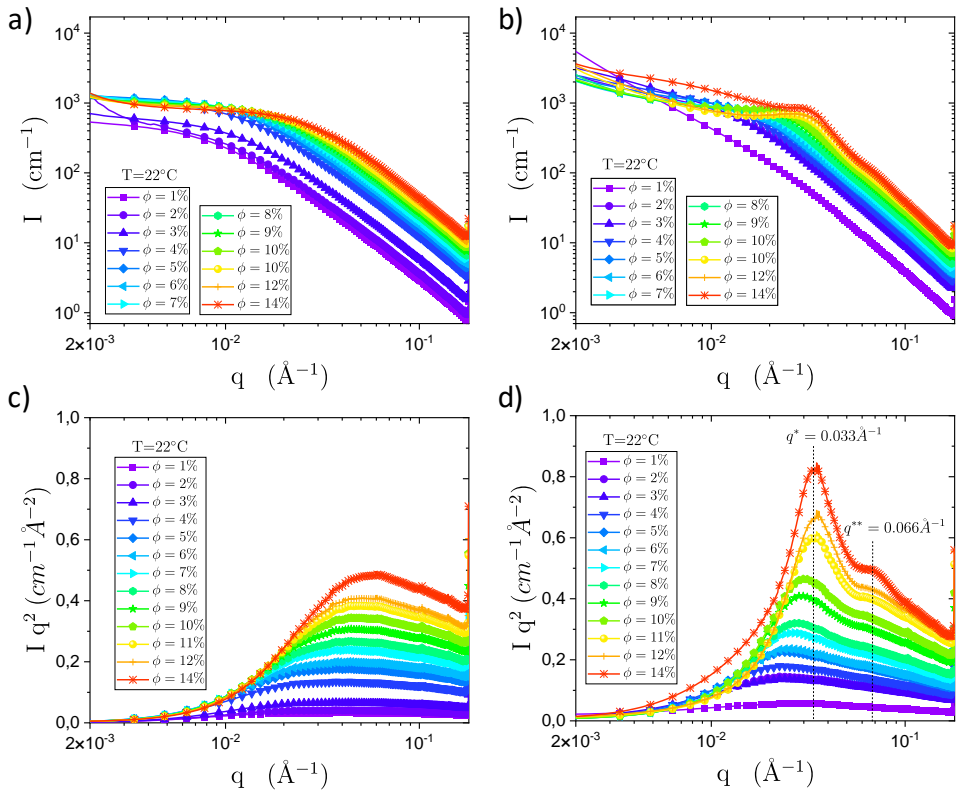

**Figure 1.** Scattering curves obtained at room temperature (i.e. T = 22 °C) as a function of the concentration ($\phi$) of bare laponite suspensions ((a) and (c)) and coated laponite suspensions ((b) and (d)). The scattering curves are shown in classical representation ((a) and (b)), *I vs. q*, and in Kratky representation ((c) and (d)), *Iq² vs. q*.

From a macroscopic point of view, the suspensions of bare laponite and of coated laponite are all in a liquid state for the concentration range under consideration. However, coated laponite suspensions could be distinguished from bare laponite suspensions by a significant increase in apparent viscosity when $\phi \geq 13$ wt.%. Moreover, we did not notice any permanent birefringence for the different suspensions of bare laponite and coated laponite (figure S6).

To summarize these first results obtained at room temperature, the addition of a copolymer layer onto the laponite surface modify the phase behavior of laponite as shown by the emergence of SAXS structural peaks at q* ~ 0.03 Å$^{-1}$ (d ~ 20 nm) and q** ~ 0.06 Å$^{-1}$ when $\phi \geq 2$ wt. %. In this concentration range, cryo-TEM observations enable to identify small stacks of laponite. In addition, all the samples remained homogeneous and liquid, with no optical birefringence detected in any of the samples (i.e. bare laponite and coated laponite) at any concentration. Altogether these results suggest that for $\phi \geq 2$ wt. %, the presence of a polymer coating enable a transition from an isotropic domain to a domain where part of the coated laponite stack to form small aggregates. These aggregates should be randomly oriented, to be compatible with the absence of birefringence, and the stacks are composed of particles separated by about 20 nm.

We then compared these results with Monte-Carlo simulations performed by Jabbari-Farouji *et al.* [11]. To facilitate comparison with these simulations, we expressed the concentrations in term of relative density, $\rho^* = D^3 \times number\ density$. Thus, when $\phi$ varies between 1 and 14 wt. %, $\rho^*$ vary from 0.14 to 2.2 for the bare nanoplatelets and from 0.44 to 6.8 for the coated nanoplatelets respectively. At first order, the phase behavior of bare laponite and polymer-coated laponite agree with the numeric prediction for a high screening parameter ($\kappa D \geq 14$). In this limit, the suspensions of bare laponite are expected to remain in the isotropic domain, whereas suspensions of coated laponite must transition from an isotropic phase to a "random stack" phase for $\rho^* \sim$ 2-3. Interestingly, the simulations enabled to derive the aggregation number using the position of the first minimum of the radial pair distribution function (g(r)) as a threshold for spatial correlations between particles. In the "random stack" domain, the predicted number of platelets per "random stack" should be around 3 in qualitative agreement with our cryo-TEM observations. We emphasize that although our observations are consistent with the numerical prediction, they remain qualitative given the difficulty of identifying nanoclusters composed of thin platelets (thickness ~ 0.9 nm) oriented perpendicular to the field of observation and at low concentration. More quantitatively, the simulations predict an inter-platelet spacing of around $0.36 \times D \sim 10$ nm, well below the distance of 20 nm obtained by SAXS measurement. We think that this difference is due to the contribution of the steric repulsion by the polymer layer around the charged platelets which prevents closer proximity.

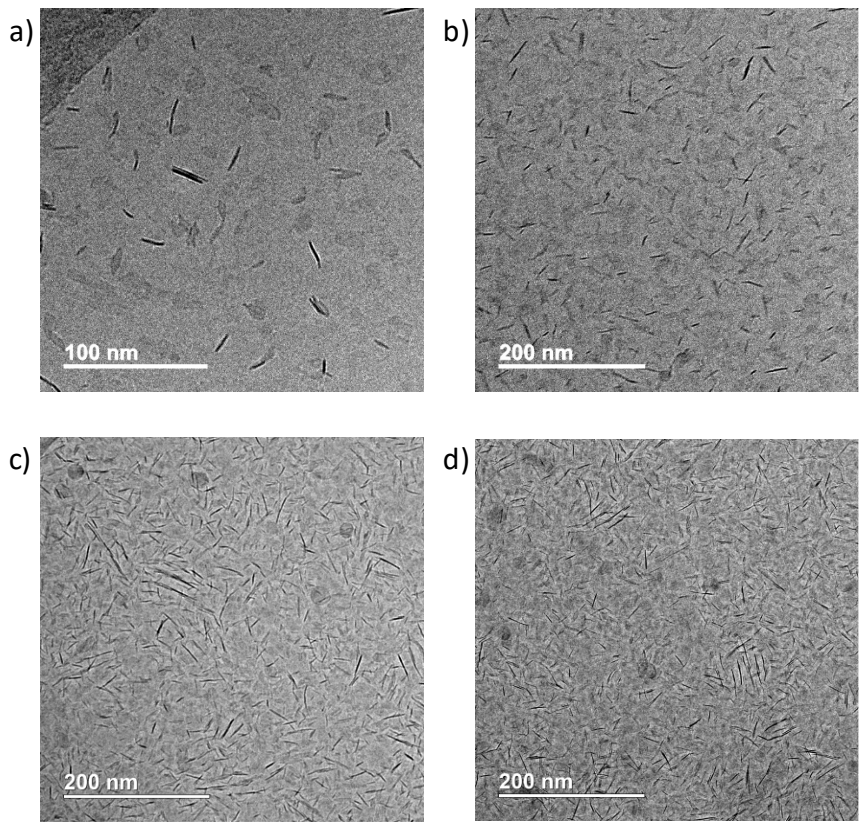

**Figure 2.** CryoTEM images of coated nanoplatelets at $\phi$ = 0.6 (a), 3 (b), 6 (c) and 9 wt.% (d). The samples were frozen from room temperature.

Overall, our results obtained at room temperature reveal a modification of the phase behavior of coated laponite suspensions which is compatible with an increase in the particle volume fraction due to the presence of a polymer shell on the surface of the laponite nanoplatelets.

**Phase behavior of bare and coated laponite at different temperatures.**

In this second part, we investigated the phase behavior of both types of nanoplatelets as function of temperature in order to probe the contribution of the thermosensitive copolymer making up the shell of coated laponite. We expect the particle volume fraction to increase with temperature due to the swelling of the copolymer shell (i.e. extension of PEO blocks in water [29]), while bare laponite particles should behave athermally. We emphasize that the work was carried out for a stoichiometric ratio of laponite and pluronic (see E.S. 2.2 and [18]) in which all pluronic chains are adsorbed onto the surface of the laponite platelets, which means that the contribution of free chains can be disregarded in the results obtained.

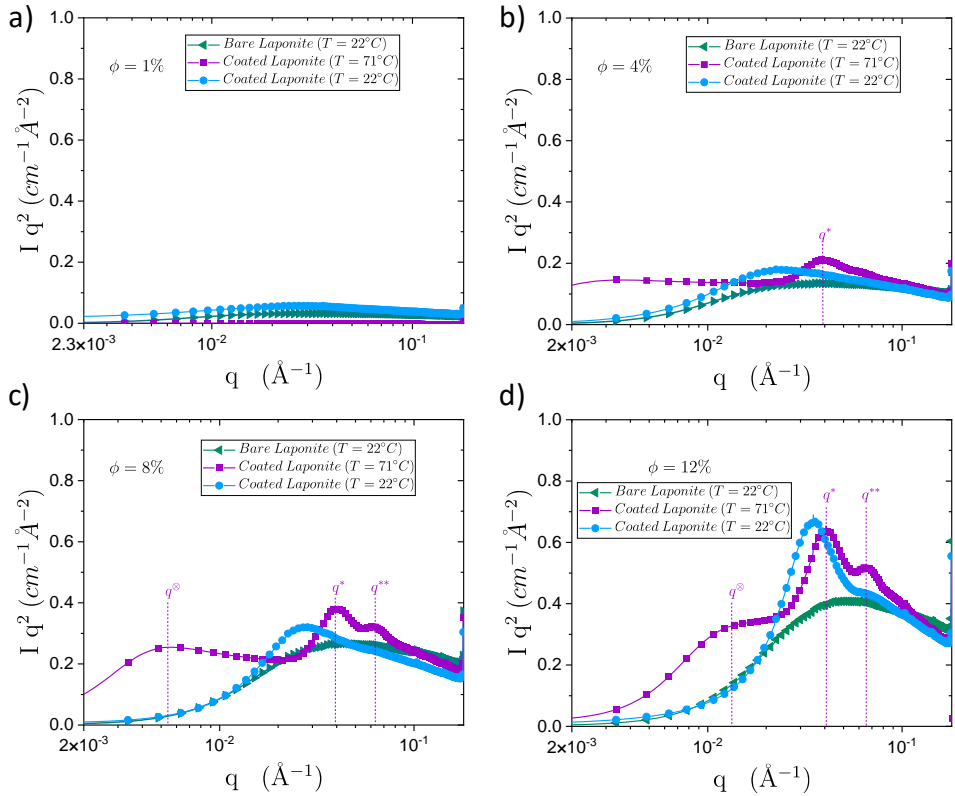

**Figure 3.** Scattering curves, in Kratky representation, for different concentrations (ϕ) in laponite: (a) 1 wt.%, (b) 4 wt.%, (c) 8 wt.% and (d) 12 wt.%. In each plot, the 3 curves correspond to measurements performed at 22 °C (bare and coated laponite) and 71 °C (coated laponite) as indicated in the inset. All the curves obtained at different concentrations and temperatures (i.e. T = 10 °C ; 22 °C ; 40 °C ; 60 °C and 71 °C) are available in classical representation (i.e. I(q)), or Kratky representation, in figures S2 and S3 of the SI respectively.

As anticipated, the scattering patterns obtained with laponite suspensions do not vary with temperature, whatever the laponite concentration (Figure S2). On the other hand, the behavior of coated laponite is temperature-dependent (Figure S3). To clearly show this effect, we have plotted the SAXS curves, in Kratky representation, for bare laponite and coated laponite supensions at two temperatures (i.e. T = 22 °C and 71 °C) for each concentration (Figures 3 and S4). Comparison of the curves obtained for coated laponite suspensions at T = 22 °C and 71 °C shows firstly that the structure peak observed at 22 °C for q = q* is also present at higher temperatures. However, for a given concentration, the peak obtained at higher temperatures is better defined (i.e. finer peak with greater amplitude) and appears at a larger q. Moreover, the harmonic peak observed for q = q** follows the same trend and is detectable at lower concentrations than for measurements at 22 °C. These effects suggest that increasing temperature favors the ordering of nanoplatelet in stacks with a decrease in inter-platelet distances.

A second effect concerns the emergence of a third structural peak when T > 40 °C, in the low q range, at $q = q^\otimes$. This peak is clearly detectable from $\phi$ = 7 wt. % and was not observed at 22°C, whatever the laponite concentration. Interestingly, the value of $q^\otimes$ and the peak amplitude increase with $\phi$ as shown in figure S5. The appearance of this low-q peak and its dependence with $\phi$ suggest that the coated laponite stacks evolve towards side-to-side positional ordering (i.e. local ordering) upon temperature increase as a result of excluded volume interactions. In this view, the distance associated with $q = q^\otimes$ would be an average inter-stack distance varying from 140 nm to 33 nm when $\phi$ vary from 7 to 14 wt.% at 71 °C.

We then studied this temperature effect at the macroscopic level by detecting the occurrence of optical birefringence and carrying out rheology measurements. From a rheological point of view, it appears that the samples behave like liquids at low laponite concentration (i.e. $\phi \leq 3$ wt.%) whatever the temperature, while a liquid/solid transition is observed at higher laponite concentration (i.e. $\phi \geq 4$ wt.%) as the temperature increases (Figures 4a and S7). At the liquid/solid transition both elastic (G') and viscous (G'') moduli show a steep increase by several decades upon reaching a temperature threshold. This thermally-activated transition is similar to that observed with micellar solutions of thermo-sensitive copolymers as a function of temperature. Interestingly, the crossover temperature, $T_{G'=G''}$, from a viscoelastic liquid (G'' > G') to a viscoelastic solid (G' > G'') decreases linearly from ~ 55 °C to ~ 40 °C when $\phi$ increases from 4 wt.% to 14 wt.% (Figure 4b).

From an optical point of view, it appears that samples in the viscoelastic solid state (i.e. $\phi \geq 4$ wt.% & T ≥ $T_{G'=G''}$) are birefringent, while samples in the liquid state (i.e. $\phi < 4$ wt.% &/or T < $T_{G'=G''}$) are non-birefringent. The observation of permanent birefringent properties is necessarily linked to a long range orientation of laponite particles (i.e. no birefringence in F127 copolymer gels) such as a nematic texture.

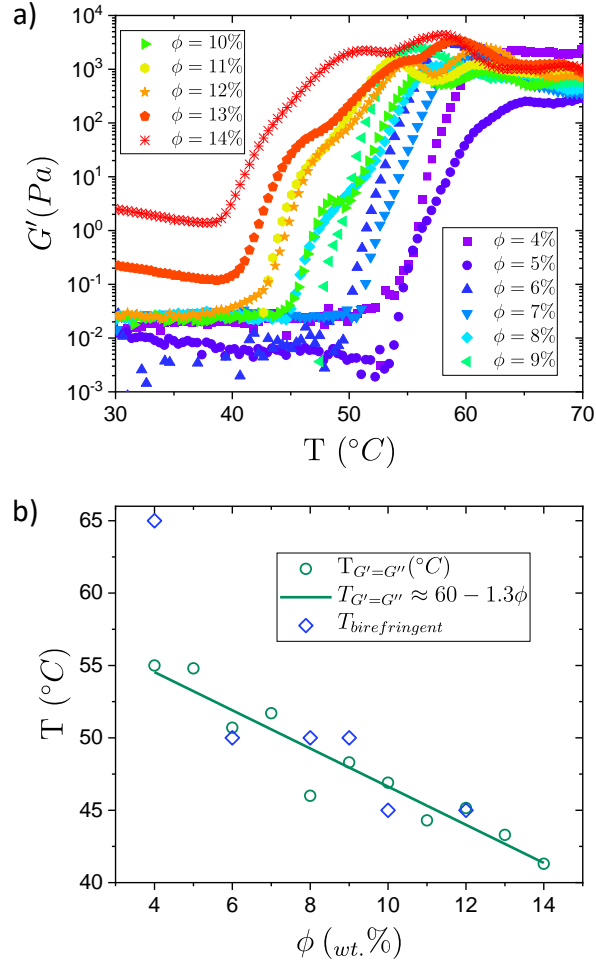

**Figure 4.** (a) Evolution of the elastic (G') moduli as a function of temperature for coated nanoplatelet suspensions at different concentrations, ϕ, ranging from 4 to 14 wt.%. (b). Evolution of the transition temperature where G' = G'' (green circles) and of the temperature at which samples become birefringent (blue diamonds) as a function of the nanoplatelet concentration (ϕ).

To summarize all these results obtained at different concentrations and temperatures, we have identified three different phases which we will refer to as: isotropic fluid, random stacks and nematic stacks, by analogy with the terminology used by Jabbari-Farouji *et al.* [11]. Our experimental approach, involving various local (i.e. SAXS and cryo-TEM) and global (i.e. rheology and optical birefringence) methods of analysis, has enabled us to identify each phases due to specific signatures:

- **Isotropic fluid**: liquid state, no birefringence, no structural peak.
- **Random stacks**: liquid state, no birefringence, 1 or 2 structural peaks ($q^*$ / $q^{**}$).
- **Nematic stacks**: solid state (G'> G''), birefringence, 2 or 3 structural peaks ($q^{\otimes}$ / $q^*$ / $q^{**}$).

Moreover, we have shown that these phases appear in peculiar concentration and temperature ranges, as illustrated in figure 5.

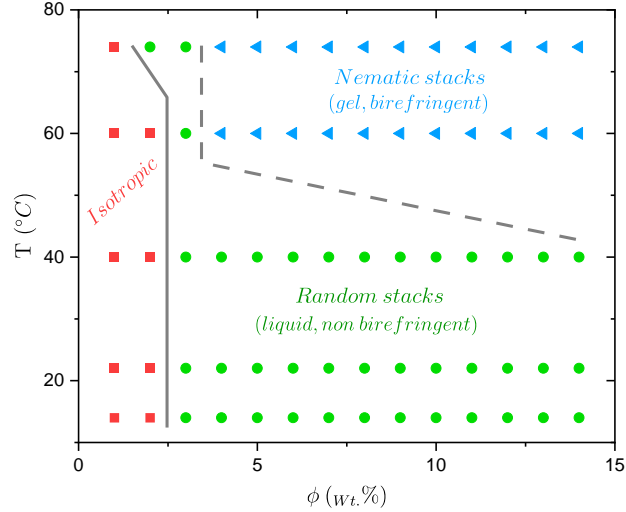

**Figure 5.** State diagram of coated nanoplatelets *versus* temperature (T) and concentration (ϕ). The dashed diagonal line separating the random stacking and the nematic stacking domains corresponds to the equation $T_{G'=G''} = 60 - 1.3\phi$ established by the rheology and the birefringence measurements in Figure 4.b."

To clarify the overall picture that emerges, the results were compared with the Monte-Carlo simulations already mentioned in the previous section to discuss the transition "Isotropic fluid → Random stacks" observed when ϕ was increased at constant temperature (i.e. T = 22 °C). In these simulations, the phase sequence "Isotropic fluid → Random stacks → Nematic stacks" was predicted following an increase in particle density ($\rho^*$) under the same screening conditions ($\kappa_D \geq$ 14) as those in which we worked. The "Nematic stacks" is the first ordered state, in terms of orientation, supposed to appear with increasing particle density [33-35]. In this state, the average size of stacks is expected to be larger than in the "Random stacks" state (i.e. the predicted maximum aggregation number vary from ~ 7 to ~ 14 platelets/stack) and no long-range positioning order is expected in any direction. This absence of long-range positioning order is in good agreement with the presence of a single low-q structural peak at q = $q^\otimes$ typical of a "liquid order" characterized by an average inter-stack distance.

In this framework, the "Random stacks (RS) → Nematic stacks (NS)" transition observed with increasing temperature when ϕ ≥ 4 wt.% suggests that the particle volume fraction increases with increasing temperature. This result is consistent with the expected thermally activated swelling of the copolymer layer deposited on the particle surface. However, when we try to quantitatively compare the numerically predicted transition thresholds ($\rho^*_{RS \to NS} \sim 4.5$) with those obtained in our study at the onset of birefringence ($\rho^*_{RS \to NS} \sim 2$), we observe differences that may be due to the polydispersity of laponite and to the coating softness, which are not taken into account in the simulations.

## Conclusion

We studied experimentally the phase behavior of thermo-responsive nanoplatelets composed of a hard core (i.e. laponite RDS) coated with a soft polymeric shell (i.e. $PEO_{100}$-$PPO_{65}$-$PEO_{100}$ copolymer) whose swelling can be controlled by temperature variation. We have identified unambiguously the specific signature associated with the different phases by combining local (i.e. SAXS and cryo-TEM) and global (i.e. rheology and optical birefringence) methods of analysis. The results are compared with the phase behavior of bare laponite nanoplatelets and with a phase diagram previously obtained numerically by another team using Monte-Carlo simulation methods. In this way, we obtained two main results. The first result is to show that the deposition of a soft layer of polymer on the laponite surface enables a phase transition from isotropic liquid towards liquid suspensions of random stacks which is not observed for bare laponite suspensions in the concentration range under our attention (i.e. ϕ ≤ 14 wt.%.) The second result is to show that an increase of nanoplatelet volume fraction triggered by temperature (i.e. swelling of the copolymer shell) could induce a phase transition from liquid suspensions of random stacks towards birefringent gels composed of nematic stacks. Both results agree with numerically predicted phase sequences expected by variation in particle density under similar charge screening conditions (i.e. κD ≥ 14), taking into account the contribution of the copolymer layer to the particle volume fraction. We believe that these results not only improve our understanding of the phase behavior of charged anisotropic colloids, but also pave the way for the control of nanoplatelet assembly by external action.

## Author Information

**Corresponding Authors**

- Imane Boucenna – Laboratoire Matière et Systèmes Complexes (MSC), UMR 7057 CNRS and Université Paris Cité ; orcid.org/ 0000-0002-7501-6686 ; Email: imane.boucenna@u-paris.fr
- Florent Carn − Laboratoire Matière et Systèmes Complexes (MSC), UMR 7057 CNRS and Université Paris Cité ; orcid.org/ 0000-0002-7842-3658 ; Email: florent.carn@u-paris.fr
- Ahmed Mourchid − Laboratoire Matière et Systèmes Complexes (MSC), UMR 7057 CNRS and Université Paris Cité ; orcid.org/ 0000-0003-2297-8504 ; Email: ahmed.mourchid@u-paris.fr

**Author Contributions:** I. B., F. C. and A. M. contributed equally.

## Data Availability

Data will be made available on request.

## Acknowledgements

We thank Dr Thomas Bizien (Soleil Synchrotron) for his assistance with carrying out SAXS experiments on Swing beamline. Soleil synchrotron is gratefully acknowledged for beam time allocation. We thank Jean-Michel Guinier (Sorbonne Université, Muséum National d'Histoire Naturelle, CNRS, IRD, Institut de Minéralogie, de Physique des Matériaux et de Cosmochimie (IMPMC, Paris, France) for the cryo-TEM observations.

## Supporting Information

The following Supporting Informations are available.
- Characterization of coated laponite by small-angle neutron scattering (SANS).
- Characterization of bare laponite (i.e. I vs q representation) by SAXS.
- Characterization of coated laponite (i.e. I vs q and $I.q^2$ vs $q^2$ plots for all conditions) by SAXS.
- Evolution of the position of the structural peak, $q^{\otimes}$, and its amplitude, $Iq^2$ vs $\phi$ at 71 °C.
- Characterization of bare and coated laponite by optical birefringence measurements at all the temperatures and concentrations.
- Rheological characterization of coated laponite suspensions.
- Photos showing the macroscopic state of the laponite suspensions after several years.
- Table summarizing the evolution of the concentration of the various components of the system and the Debye length as a function of the concentration of Laponite.

# Supporting Information

**Phase Behavior of Thermo-Responsive Nanoplatelets**.

**Imane Boucenna,* Florent Carn* and Ahmed Mourchid***

Matière et Systèmes Complexes (MSC), UMR 7057 CNRS and Université Paris Cité

75205 Paris Cedex 13, France

*Corresponding authors:
- Imane Boucenna ; Email: imane.boucenna@u-paris.fr
- Florent Carn ; Email: florent.carn@u-paris.fr
- Ahmed Mourchid ; Email: ahmed.mourchid@u-paris.fr

______________________________________

Number of pages: 12
Number of figures: 9
Number of schemes: 0
Number of tables: 1

______________________________________

## Table of Content

# 1. Characterization of coated laponite by small-angle neutron scattering (SANS).

The neutron scattering experiments were carried out at Léon Brillouin Laboratory (CNRS-CEA, Saclay, France) on PAXY beamline. Two experimental configurations were chosen to cover scattering wavevectors q ranging from 0.003 and 0.3 Å$^{-1}$ by choosing incident neutron wavelengths of 5 and 10 Å and sample-2 dimensional detector distances of 2 and 5 m, in first and second configuration respectively. We used quartz cells of 2 mm path length, for dilute samples in deuterated water. They were filled at 4°C and preliminarily equilibrated at the measuring temperature before being transferred to the cell holder. The scattering from the solvent in the quartz cell was subtracted off from the data which were normalized by the incoherent scattering signal of H$_2$O to yield absolute scattering intensities in cm$^{-1}$. The neutron scattering length densities of coating polymer, nanoparticles and solvent are: $\rho_{pol}$ = 0.052 10$^{-5}$ Å$^{-2}$, $\rho_{lap}$ = 0.418 10$^{-5}$ Å$^{-2}$ and $\rho_{D2O}$ = 0.633 10-5 Å$^{-2}$. The contrast between the copolymer and D$_2$O solvent, $(\rho_{pol} - \rho_{D2O})^2$, is equal to 33.76 10$^{-12}$ Å$^{-4}$, while for the laponite in D$_2$O the contrast is $(\rho_{lap} - \rho_{D2O})^2$ 4.62 10-12 Å$^{-4}$. The SANS data on bare NPs were modeled by using iterative least-squares methods and the appropriate form factor of polydisperse hard thin disks coated with a thin layer of copolymer chains.

The form factor of monodisperse discs is $P(q) = \int_0^{\frac{\pi}{2}} \sin\theta \, d\theta \, [V_{lap} \, \Delta\rho \, K(q, l, r, \theta)]^2$

Where $K(q, l, r) = \frac{\sin[ql(\cos\theta)/2]}{ql(\cos\theta)/2} \frac{2J_1[qr \sin\theta/2]}{qr \sin\theta}$; l and r are particle thickness and radius respectively.

In dilute suspensions the structure factor is assumed to be 1 and by taking into account nanoparticle radius polydispersity, using log-normal distribution of radii, n(r$_0$, r, σ):

$$I(q) = \frac{Nanoparticle \; volume \; fraction}{V_{lap}} \int_{r=0}^{\infty} n(r_0, r, \sigma) P(q) \, dr$$

and $n(r_0, r, \sigma) = \frac{\exp[-0.5(\frac{\ln r_0 - \ln r}{\sigma} - 0.5 \sigma)^2]}{(2\pi)^{0.5} \, r \, \sigma}$, where r$_0$ is mean radius and σ is the polydispersity.

The form factor of coated discs is:

$$P(q) = \int_0^{\frac{\pi}{2}} \sin\theta \, d\theta \{(V_1 \Delta\rho_1) K(q, l + 2\Delta e, r + \Delta e, \theta) + (V_2 \Delta\rho_2) K(q, l, r, \theta)\}^2$$

Where $V_1 = \pi(r_p+\Delta e)^2(l+2\Delta e)$; $\Delta\rho_1=(\rho_{pol}- \rho_{D2O})$; $V_2 = \pi(r_p)^2 l$; $\Delta\rho_2=[(\rho_{lap}- \rho_{D2O})- (\rho_{pol}- \rho_{D2O})] =(\rho_{lap}- \rho_{pol})$. For fitting coated nanoplatelet form factor, we used fixed mean radius and polydispersity values previously determined from fitting the form factor of bare laponite. The same procedure was used to fit the form factor from SAXS data.

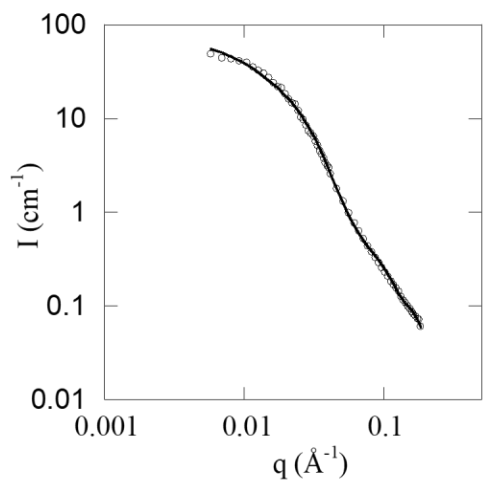

**Figure S1**. Small-angle neutron scattering intensity measured at ambient temperature for coated Laponite at 2 wt. % (○). The continuous line correspond to a fit using the form factor of coated polydisperse disks, with fixed diameter, thickness and (LogNormal) polydispersity (280 Å, 9.2 Å and 0.30 respectively), and a varying coating extension which is found to be $\Delta e = 32$ Å.

## 2. Characterization of bare laponite by small-angle X-ray scattering (SAXS).

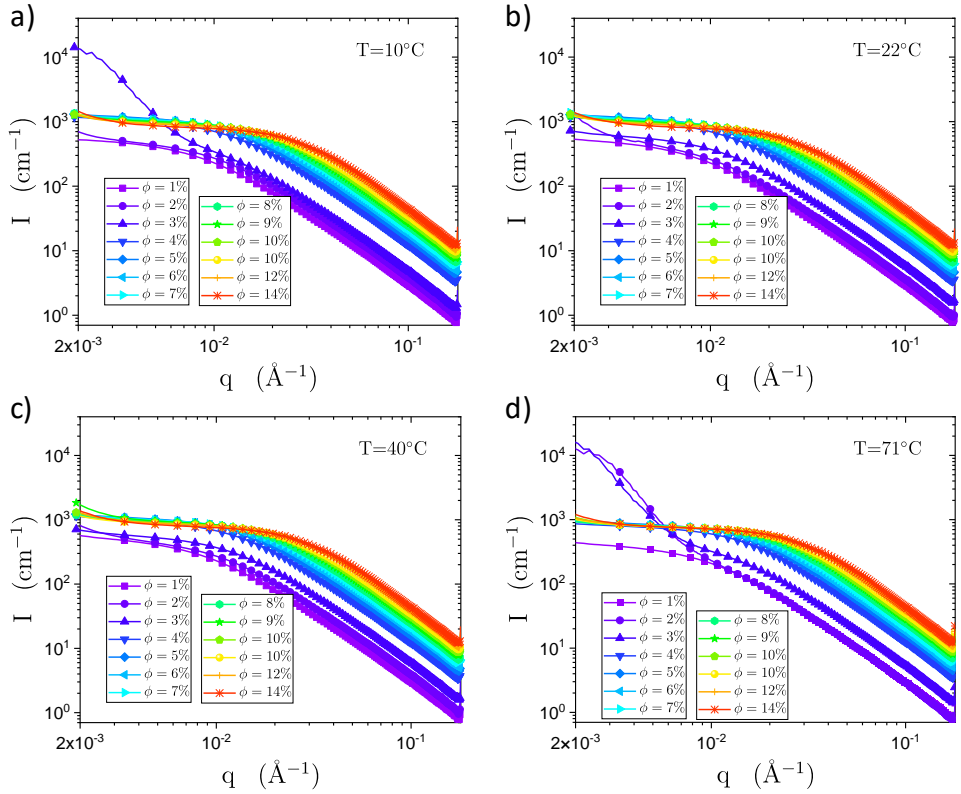

**Figure S2**. Scattering curves obtained at different temperature (i.e. T = 10°C ; 22°C ; 40°C and 71°C), indicated in the inset, as a function of the bare laponite mass fraction (ϕ). The continuous lines are guide for the eyes. We underline that the slight upturn observed on 3 points at very low q for 3 curves (i.e. ϕ= 3% at 10°C and 71°C ; ϕ= 2% at 71°C) correspond to punctual problems of SAXS acquisition/treatment (i.e., sample heterogeneity, subtraction issues, micro-scratches, capillary orientation, etc.) which will not be taken into account in our data analysis.

## 3. Characterization of coated laponite by small-angle X-ray scattering (SAXS).

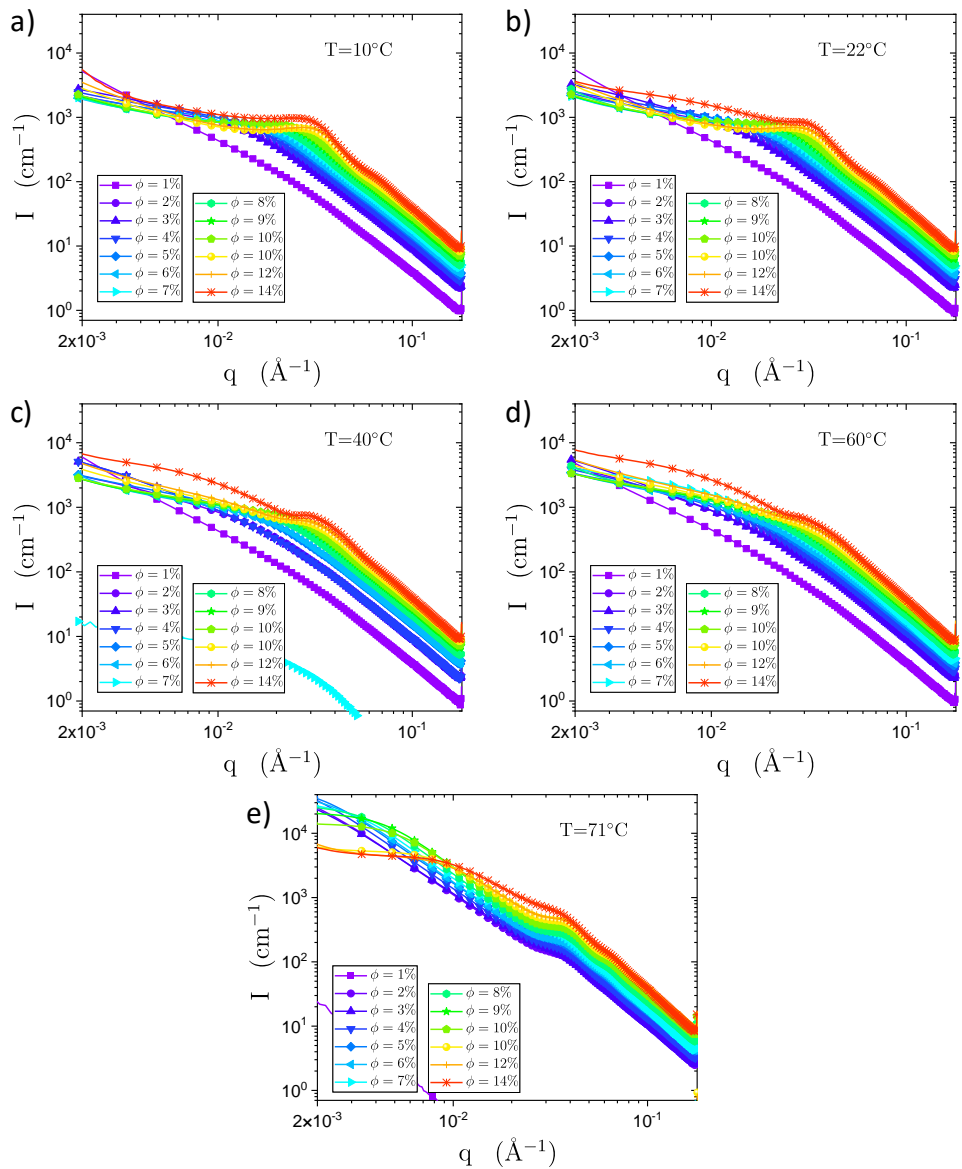

**Figure S3**. Scattering curves obtained for different coated laponite mass fraction (ϕ), indicated in the inset, and for different temperatures: a) 10°C ; b) 22°C ; c) 40°C ; d) 60°C and e) 71°C. The continuous lines are guide for the eyes.

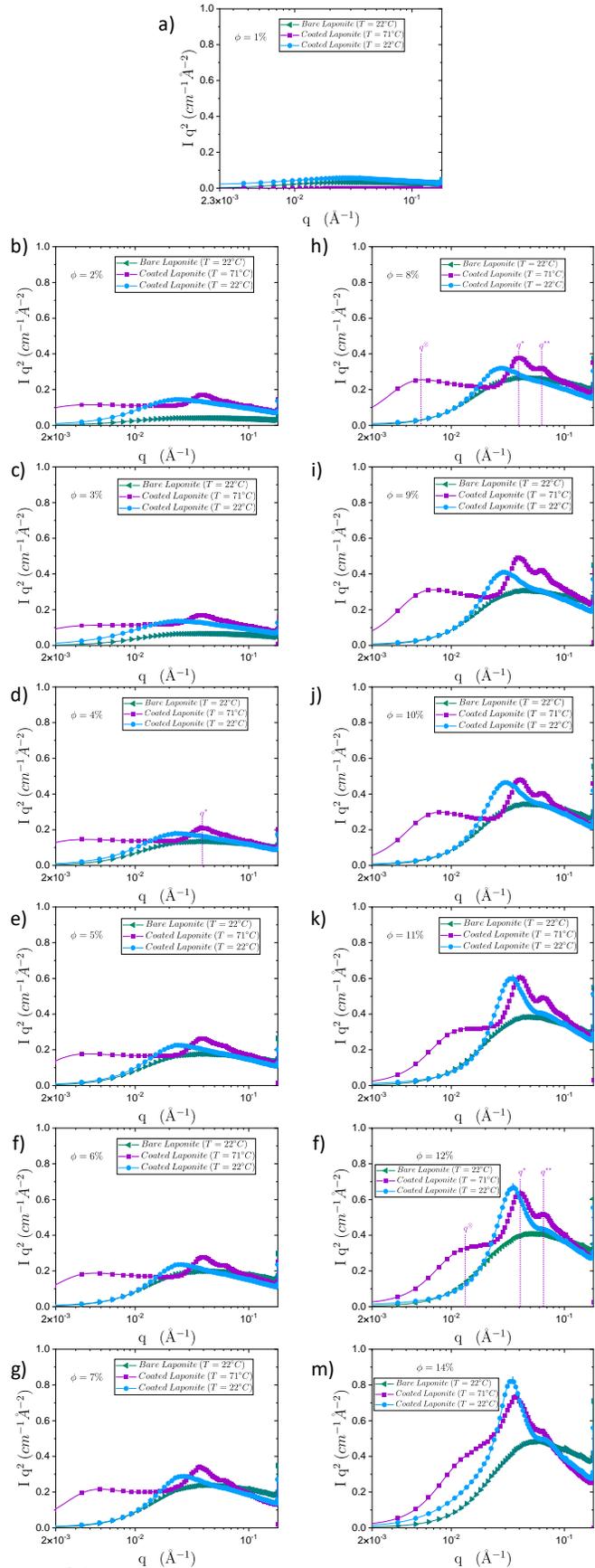

**Figure S4**. Scattering curves, in Kratky representation (i.e. Iq² vs. q), obtained at different temperatures (i.e. T = 10°C ; 22°C ; 40°C ; 60°C and 71°C), indicated in the inset, as a function of the laponite mass fraction (ϕ) for bare laponite suspensions (left column) and coated laponite suspensions (right column): a) Φ = 1% ; b) Φ = 2% ; c) Φ = 3% ; d) Φ = 4% ; e) Φ = 5% ; f) Φ = 6% ; g) Φ = 7% ; h) Φ = 8% ; i) Φ = 9% ; j) Φ = 10% ; k) Φ = 11% ; l) Φ = 12% ; m) Φ = 14%. The continuous lines are guide for the eyes.

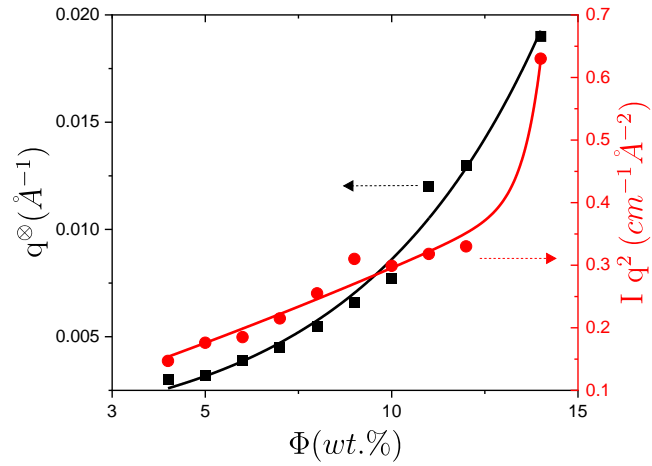

**Figure S5**. Evolution of the position of the structural peak, $q^{\otimes}$ (black square), and its amplitude, $Iq^2$ (red disc) as a function of coated laponite concentration at 71 °C. Solid lines are guides for the eyes.

## 4. Characterization of bare and coated laponite by optical birefringence measurements.

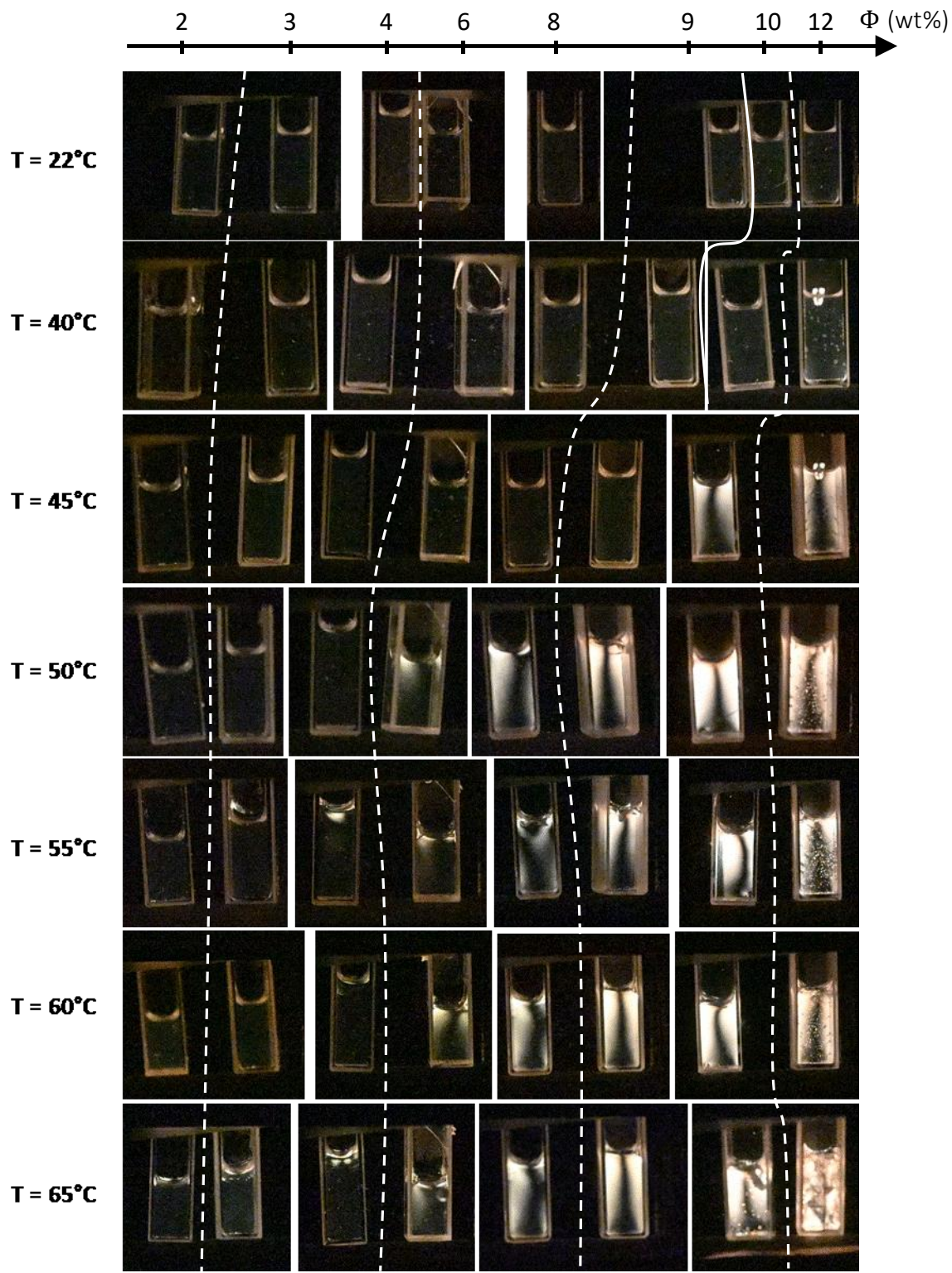

**Figure S6.** Coated laponite at 2, 3, 4, 6, 8, 9, 10 and 12 wt. % (from left to right observed between cross-polarizers at 22, 40, 45, 50, 55, 60 and 65°C from top to bottom).

## 5. Rheological characterization of coated laponite suspensions.

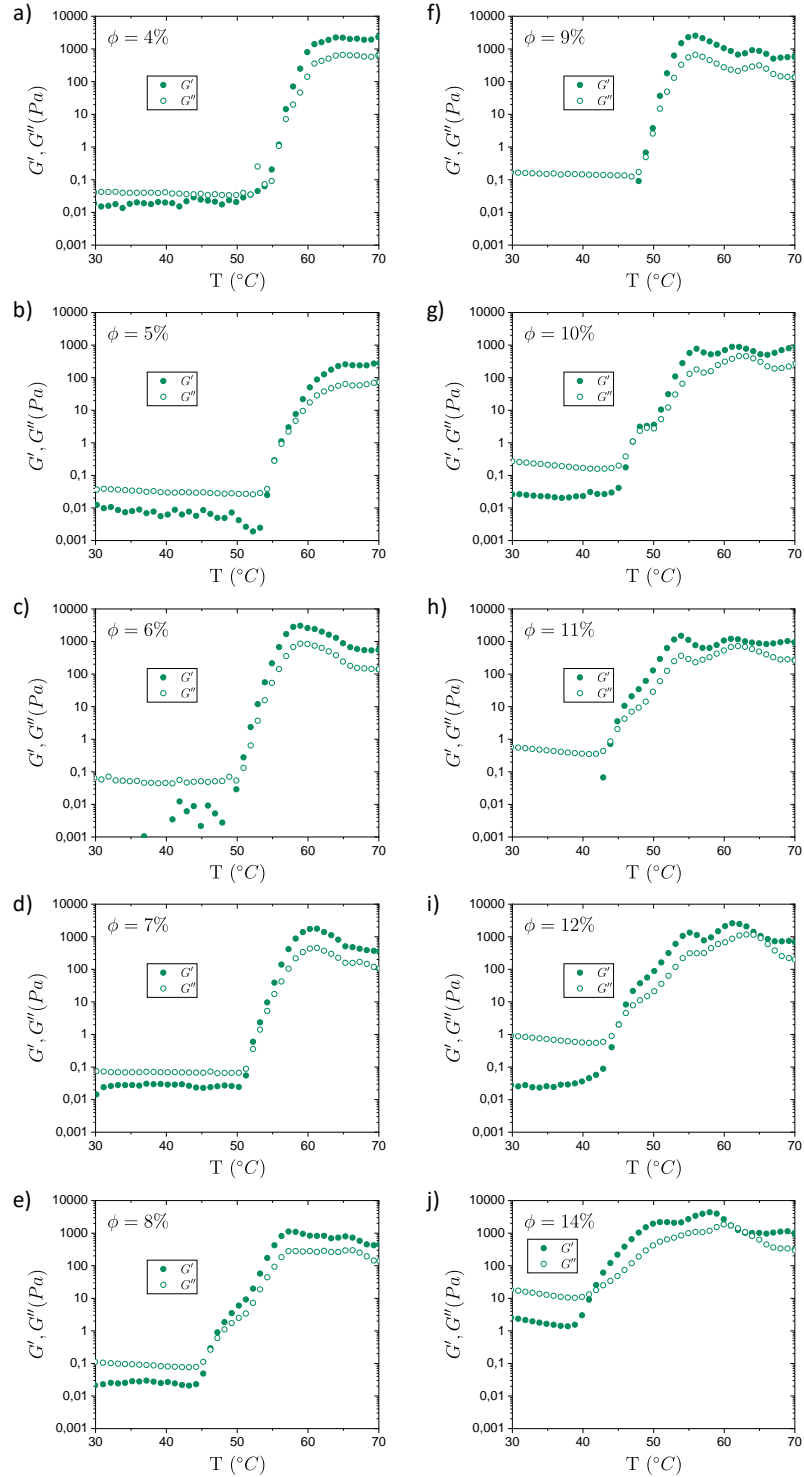

**Figure S7**. Evolution of the elastic (G', full symbol) and loss (G'', empty symbol) moduli as a function of temperature for coated nanoplatelet suspensions at different mass concentration ranging from 4 $_{wt}$% to 14 $_{wt}$% as indicated in the inset: a) $\Phi = 4\%$ ; b) $\Phi = 5\%$ ; c) $\Phi = 6\%$ ; d) $\Phi = 7\%$ ; e) $\Phi = 8\%$ ; f) $\Phi = 9\%$ ; g) $\Phi = 10\%$ ; h) $\Phi = 11\%$ ; i) $\Phi = 12\%$ ; j) $\Phi = 14\%$.

6. **Macroscopic state of the laponite suspensions after several years.**

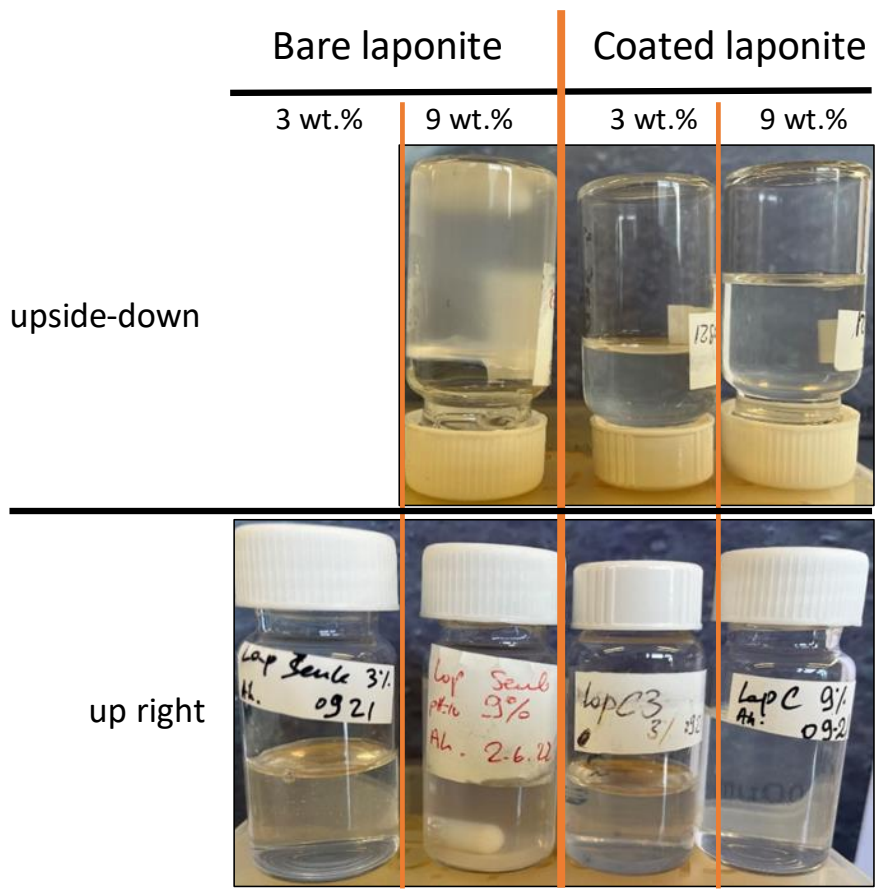

**Figure S8**. Photos taken at room temperature of bare laponite and coated laponite suspensions after being stored for several years at 4°C. The most recent sample dates back to June 2022 (i.e. bare laponite at 9 wt.%), while the oldest dates back to September 2021 (i.e. bare laponite at 3 wt.% and the two samples of coated laponite). The 3 samples in the upper image are turned upside-down

## 7. Estimation of ionic strength and Debye length.

The samples were prepared by mixing the solid laponite powder with aqueous solutions at fixed pH = 10 with NaOH. The nanoplatelets bear both negative charges on the faces and positive charges on the rims. The number of negative charges per nanoplatelet have been measured for different concentrations and found to be about 750 e. The number of the rim charges is more difficult to measure, however the typical estimate is roughly 10% of the negative charges [B. Ruzicka, and E. Zaccarelli, A fresh look at the Laponite phase diagram. Soft Matter, 2011, 7, 1268]. We used the value of 75 rim charges per nanoplatelet to determine the concentration of excess ($4Na^+$, $P_2O_7^{4-}$) salt in each solution knowing that 75 pyrophosphate anions are attached to each nanoplatelet and the remaining acts as added salt that constitutes the double layer. We considered a total pyrophosphate concentration of 8.2 g per 100 g of dry laponite powder, as indicated by the supplier. The ionic strength, $I = \frac{1}{2}\sum_i c_i Z_i^2$, and thus the Debye length, $\kappa^{-1}$, of the suspensions could be estimated by taking into account the contribution of $Na_4P_2O_7$ (i.e. $4Na^+$, $P_2O_7^{4-}$) in excess (i.e. that don't interact with the positive charges on the rims). Thus, we assume that $Na^+$ counter ions of the laponite platelets do not contribute significantly to the ionic strength. [A. Mourchid et al., Langmuir 1995, 11, 1942 ; M. Dubois et al. J. Chem. Phys. 1992, 96, 2278]. In our experiments, we estimate that the variation in laponite mass fraction ($\phi$) from 1% to 14% is associated with a variation in Debye length (i.e. $\kappa^{-1} = \sqrt{\left(\frac{\varepsilon_r \varepsilon_0 kT}{2e^2} \times \frac{1}{I}\right)}$ with I the ionic strength calculated by only taking into account the $Na_4P_2O_7$ in excess) from approximately 2.3 to 0.6 nm (Table S9).

The concentration of each species (i.e. nanoplatelet, copolymer, total pyrophosphate present in the raw material and calculated free pyrophosphate in each solution) are given in the table S9 with the calculated ionic strength, I, and Debye length, $\kappa^{-1}$:

| Nanoplatelets (wt. %) | Polymer (wt. %) | Anions (M) | Rim charges (M) | Free anions* (M) | Ionic strength** (M) | Debye length*** (Å) |
|---|---|---|---|---|---|---|
| 1  | 0.81  | 0.0031 | 0.0010 | 0.0021 | 0.017 | 22.7 |
| 2  | 1.62  | 0.0062 | 0.0021 | 0.0041 | 0.035 | 16.0 |
| 3  | 2.43  | 0.0093 | 0.0031 | 0.0062 | 0.052 | 13.1 |
| 4  | 3.24  | 0.0124 | 0.0042 | 0.0082 | 0.070 | 11.3 |
| 5  | 4.05  | 0.0155 | 0.0052 | 0.0103 | 0.087 | 10.1 |
| 6  | 4.86  | 0.0186 | 0.0062 | 0.0123 | 0.105 | 9.3  |
| 7  | 5.67  | 0.0217 | 0.0073 | 0.0144 | 0.122 | 8.6  |
| 8  | 6.48  | 0.0248 | 0.0083 | 0.0165 | 0.140 | 8.0  |
| 9  | 7.29  | 0.0279 | 0.0094 | 0.0185 | 0.157 | 7.6  |
| 10 | 8.1   | 0.0310 | 0.0104 | 0.0206 | 0.175 | 7.2  |
| 11 | 8.91  | 0.0341 | 0.0114 | 0.0226 | 0.192 | 6.8  |
| 12 | 9.72  | 0.0372 | 0.0125 | 0.0247 | 0.210 | 6.6  |
| 13 | 10.53 | 0.0403 | 0.0135 | 0.0268 | 0.227 | 6.3  |
| 14 | 11.34 | 0.0434 | 0.0146 | 0.0288 | 0.245 | 6.1  |

**Table S9.** Table summarizing the evolution of the concentration of the various components of the system and of the Debye length as a function of the concentration of laponite.

\* We assume that the free anions are only related to the contribution of the $Na_4P_2O_7$ (i.e. $4 Na^+ + P_2O_7^{4-}$) in excess.
\*\* I = 0.5\*([ $Na^+$]\*1+[ $P_2O_7^{4-}$]\*4²) = 8.5 \* [$Na_4P_2O_7$]$_{excess}$= 8.5 \* [Free anions].
\*\*\* $\kappa^{-1} \approx \frac{0.304}{\sqrt{I}}$ in water at 20°C.

We also measured at 20°C the evolution of the conductivity of the coated Laponite suspension as a function of the Laponite concentration as shown in figure S10. The measurement were done with a NanoZS apparatus (Malvern Instrument) using a standard folded capillary zeta cell. The observed variation of conductivity qualitatively reflects the expected effect of the ionic strength.

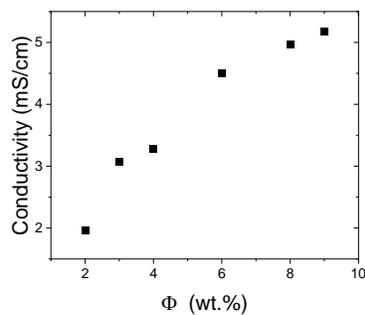

**Figure S10.** Evolution of the conductivity of the coated laponite suspensions as a function of the laponite concentration (T = 20°C).

# Supporting Information

**Phase Behavior of Thermo-Responsive Nanoplatelets**.

**Imane Boucenna,* Florent Carn* and Ahmed Mourchid***

Matière et Systèmes Complexes (MSC), UMR 7057 CNRS and Université Paris Cité

75205 Paris Cedex 13, France

*Corresponding Authors
- Imane Boucenna ; Email: imane.boucenna@u-paris.fr
- Florent Carn ; Email: florent.carn@u-paris.fr
- Ahmed Mourchid ; Email: ahmed.mourchid@u-paris.fr

─────────────────────────────────────

Number of pages: 12
Number of figures: 9
Number of schemes: 0
Number of tables: 1

─────────────────────────────────────

## Table of Content

## 8. Characterization of coated laponite by small-angle neutron scattering (SANS).

The neutron scattering experiments were carried out at Léon Brillouin Laboratory (CNRS-CEA, Saclay, France) on PAXY beamline. Two experimental configurations were chosen to cover scattering wavevectors q ranging from 0.003 and 0.3 Å$^{-1}$ by choosing incident neutron wavelengths of 5 and 10 Å and sample-2 dimensional detector distances of 2 and 5 m, in first and second configuration respectively. We used quartz cells of 2 mm path length, for dilute samples in deuterated water. They were filled at 4°C and preliminarily equilibrated at the measuring temperature before being transferred to the cell holder. The scattering from the solvent in the quartz cell was subtracted off from the data which were normalized by the incoherent scattering signal of H$_2$O to yield absolute scattering intensities in cm$^{-1}$. The neutron scattering length densities of coating polymer, nanoparticles and solvent are: $\rho_{pol}$ = 0.052 10$^{-5}$ Å$^{-2}$, $\rho_{lap}$ = 0.418 10$^{-5}$ Å$^{-2}$ and $\rho_{D2O}$ = 0.633 10-5 Å$^{-2}$. The contrast between the copolymer and D$_2$O solvent, $(\rho_{pol} - \rho_{D2O})^2$, is equal to 33.76 10$^{-12}$ Å$^{-4}$, while for the laponite in D$_2$O the contrast is $(\rho_{lap} - \rho_{D2O})^2$ 4.62 10-12 Å$^{-4}$. The SANS data on bare NPs were modeled by using iterative least-squares methods and the appropriate form factor of polydisperse hard thin disks coated with a thin layer of copolymer chains.

The form factor of monodisperse discs is $P(q) = \int_0^{\frac{\pi}{2}} \sin\theta \, d\theta \, [V_{lap} \, \Delta\rho \, K(q,l,r,\theta)]^2$

Where $K(q,l,r) = \frac{\sin[ql(\cos\theta)/2]}{ql(\cos\theta)/2} \frac{2J_1[qr \sin\theta/2]}{qr \sin\theta}$; l and r are particle thickness and radius respectively. In dilute suspensions the structure factor is assumed to be 1 and by taking into account nanoparticle radius polydispersity, using log-normal distribution of radii, n(r$_0$, r, σ):

$$I(q) = \frac{Nanoparticle \; volume \; fraction}{V_{lap}} \int_{r=0}^{\infty} n(r_0, r, \sigma) P(q) \, dr$$

and $n(r_0, r, \sigma) = \frac{\exp[-0.5(\frac{\ln r_0 - \ln r}{\sigma} - 0.5\sigma)^2]}{(2\pi)^{0.5} \, r \, \sigma}$, where r$_0$ is mean radius and σ is the polydispersity.

The form factor of coated discs is:

$$P(q) = \int_0^{\frac{\pi}{2}} \sin\theta \, d\theta \{(V_1 \Delta\rho_1) K(q, l+2\Delta e, r+\Delta e, \theta) + (V_2 \Delta\rho_2) K(q, l, r, \theta)\}^2$$

Where $V_1 = \pi(r_p+\Delta e)^2(l+2\Delta e)$; $\Delta\rho_1=(\rho_{pol}- \rho_{D2O})$; $V_2 = \pi(r_p)^2 l$; $\Delta\rho_2=[(\rho_{lap}- \rho_{D2O})- (\rho_{pol}- \rho_{D2O})]$ $=(\rho_{lap}- \rho_{pol})$. For fitting coated nanoplatelet form factor, we used fixed mean radius and polydispersity values previously determined from fitting the form factor of bare laponite. The same procedure was used to fit the form factor from SAXS data.

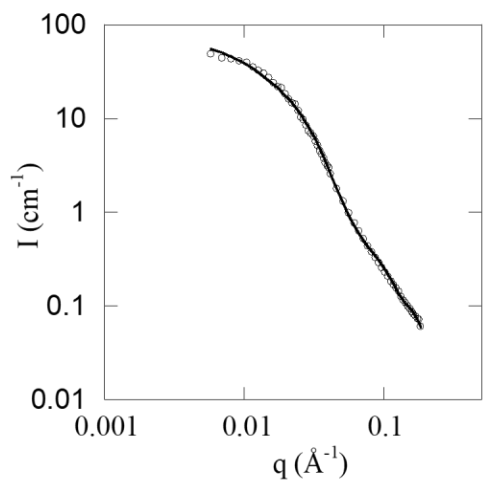

**Figure S1**. Small-angle neutron scattering intensity measured at ambient temperature for coated Laponite at 2 wt. % (○). The continuous line correspond to a fit using the form factor of coated polydisperse disks, with fixed diameter, thickness and (LogNormal) polydispersity (280 Å, 9.2 Å and 0.30 respectively), and a varying coating extension which is found to be $\Delta e = 32$ Å.

## 9. Characterization of bare laponite by small-angle X-ray scattering (SAXS).

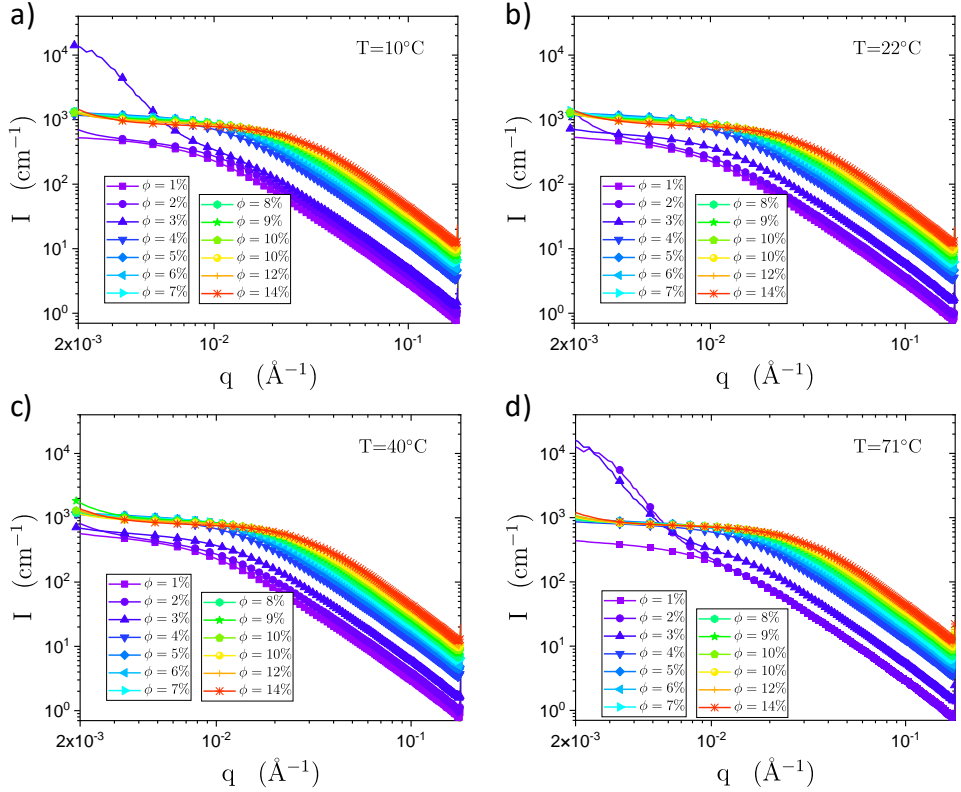

**Figure S2**. Scattering curves obtained at different temperature (i.e. T = 10°C ; 22°C ; 40°C and 71°C), indicated in the inset, as a function of the bare laponite mass fraction (ϕ). The continuous lines are guide for the eyes. We underline that the slight upturn observed on 3 points at very low q for 3 curves (i.e. ϕ= 3% at 10°C and 71°C ; ϕ= 2% at 71°C) correspond to punctual problems of SAXS acquisition/treatment (i.e., sample heterogeneity, subtraction issues, micro-scratches, capillary orientation, etc.) which will not be taken into account in our data analysis.

## 10. Characterization of coated laponite by small-angle X-ray scattering (SAXS).

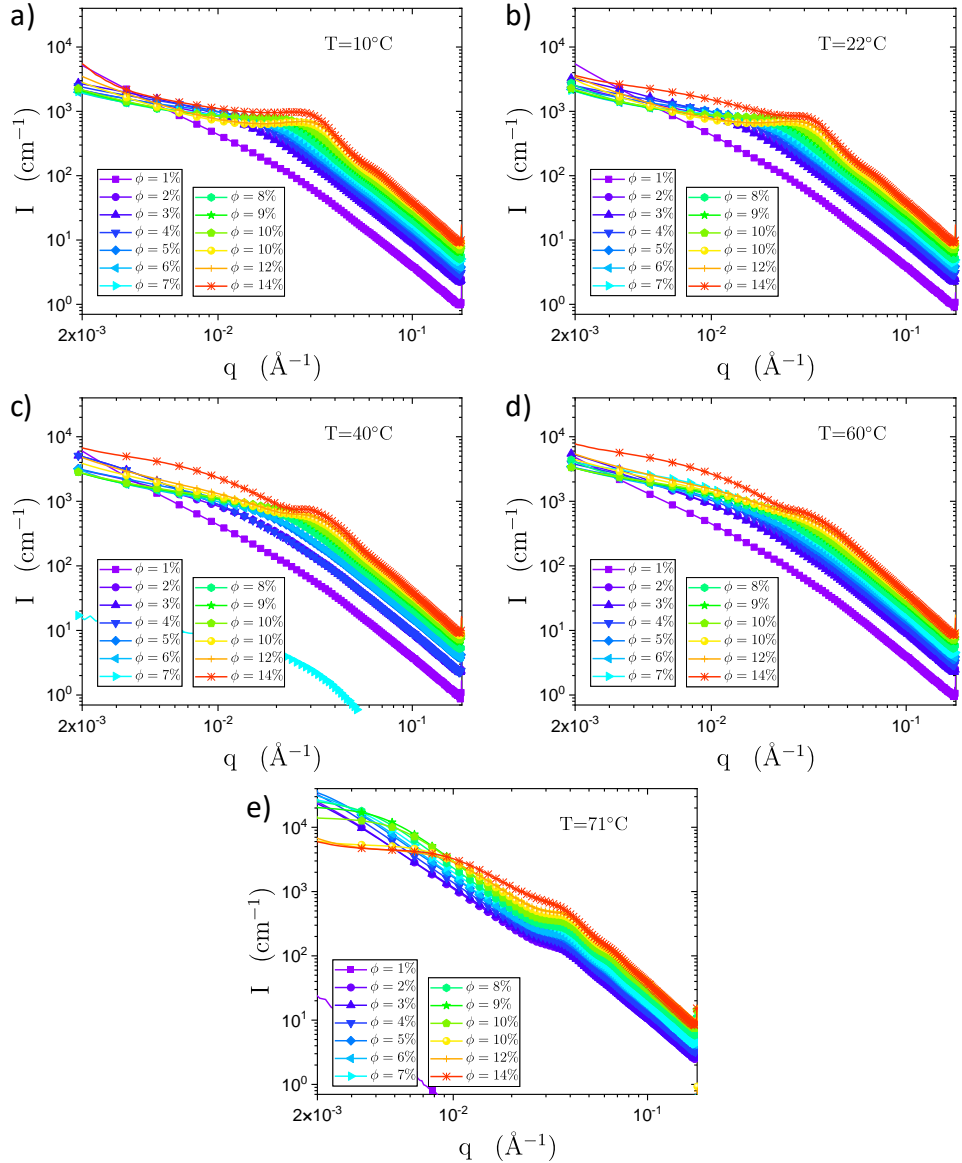

**Figure S3**. Scattering curves obtained for different coated laponite mass fraction (ϕ), indicated in the inset, and for different temperatures: a) 10°C ; b) 22°C ; c) 40°C ; d) 60°C and e) 71°C. The continuous lines are guide for the eyes.

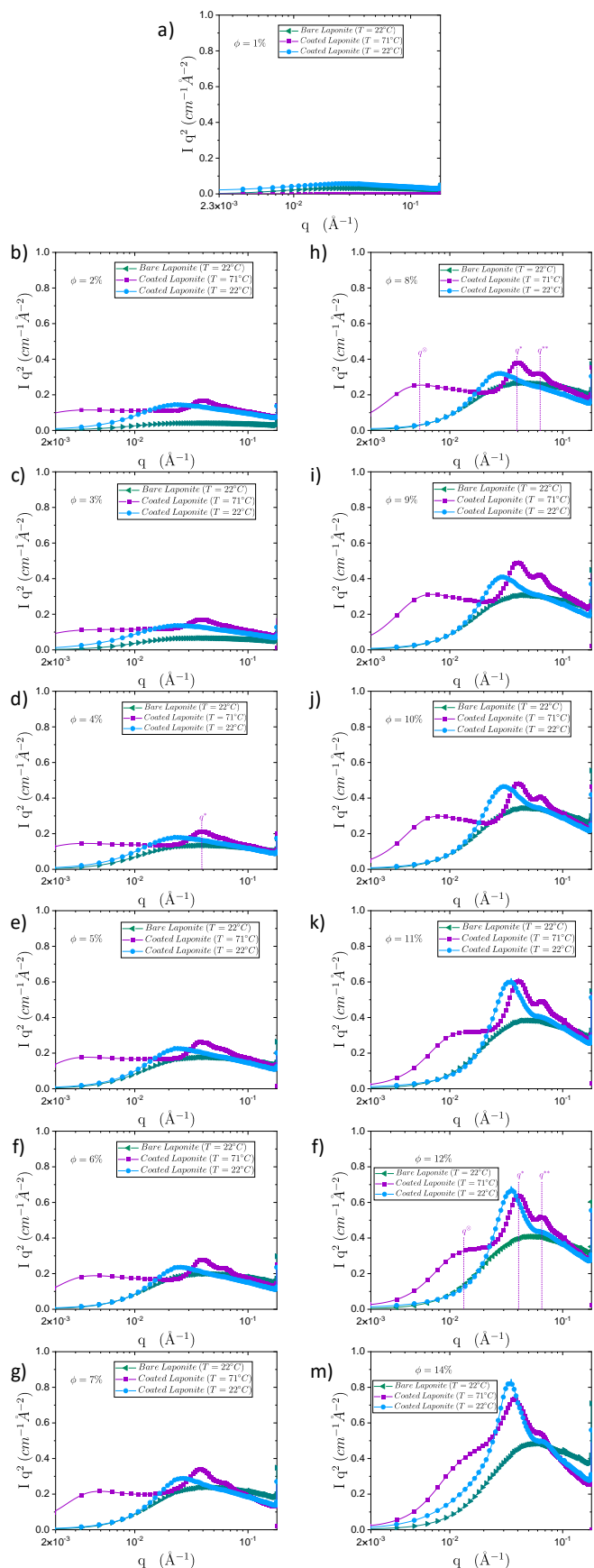

**Figure S4**. Scattering curves, in Kratky representation (i.e. Iq² vs. q), obtained at different temperatures (i.e. T = 10°C ; 22°C ; 40°C ; 60°C and 71°C), indicated in the inset, as a function of the laponite mass fraction (ϕ) for bare laponite suspensions (left column) and coated laponite suspensions (right column): a) Φ = 1% ; b) Φ = 2% ; c) Φ = 3% ; d) Φ = 4% ; e) Φ = 5% ; f) Φ = 6% ; g) Φ = 7% ; h) Φ = 8% ; i) Φ = 9% ; j) Φ = 10% ; k) Φ = 11% ; l) Φ = 12% ; m) Φ = 14%. The continuous lines are guide for the eyes.

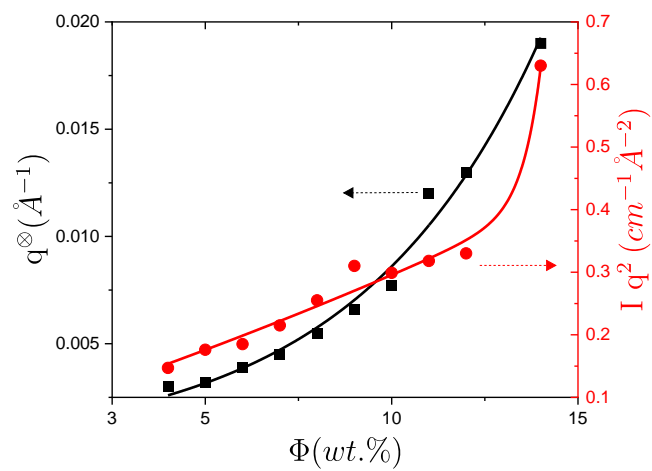

**Figure S5**. Evolution of the position of the structural peak, $q^{\otimes}$ (black square), and its amplitude, $Iq^2$ (red disc) as a function of coated laponite concentration at 71 °C. Solid lines are guides for the eyes.

## 11. Characterization of bare and coated laponite by optical birefringence measurements.

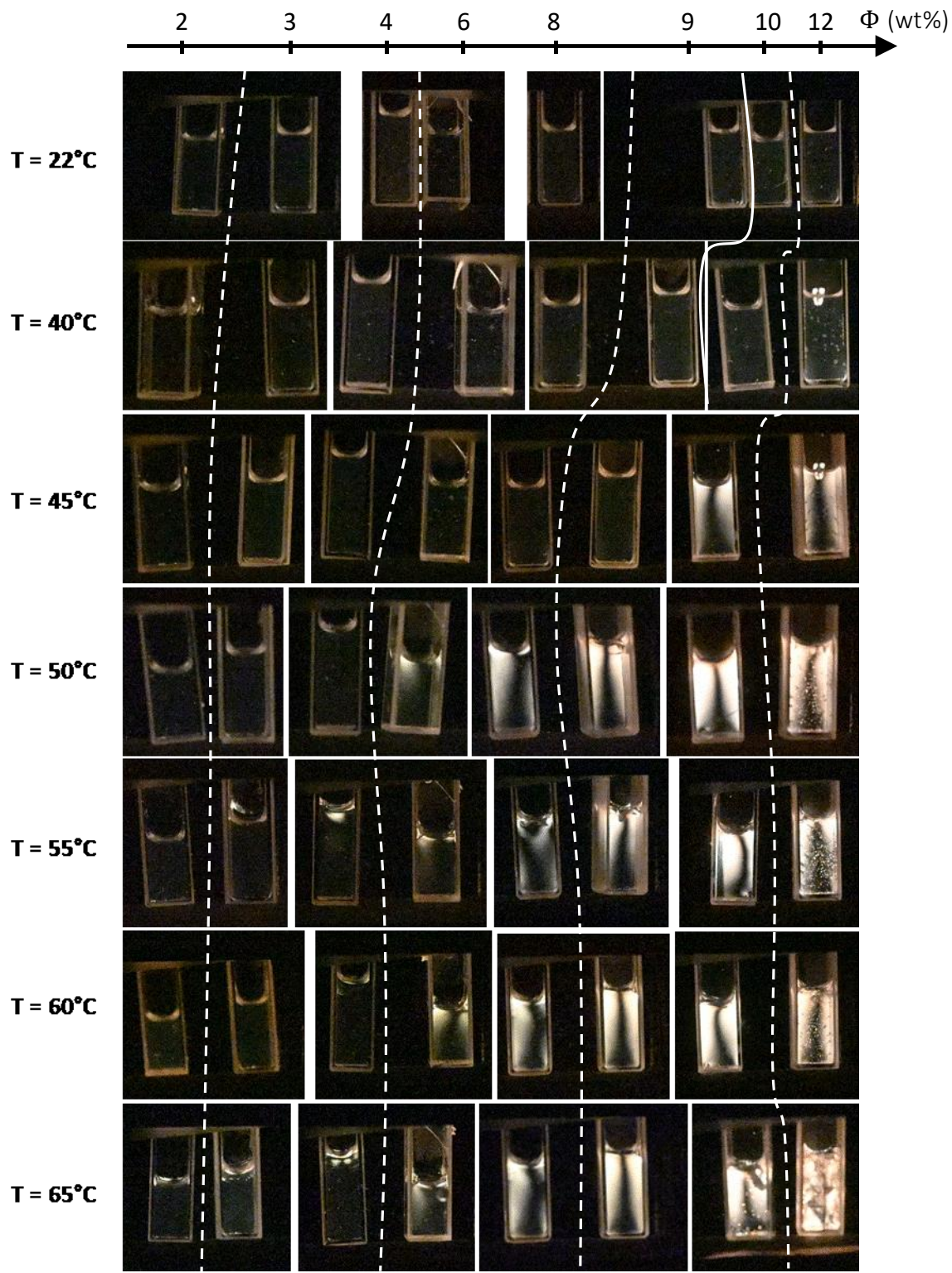

**Figure S6.** Coated laponite at 2, 3, 4, 6, 8, 9, 10 and 12 wt. % (from left to right observed between cross-polarizers at 22, 40, 45, 50, 55, 60 and 65°C from top to bottom).

## 12. Rheological characterization of coated laponite suspensions.

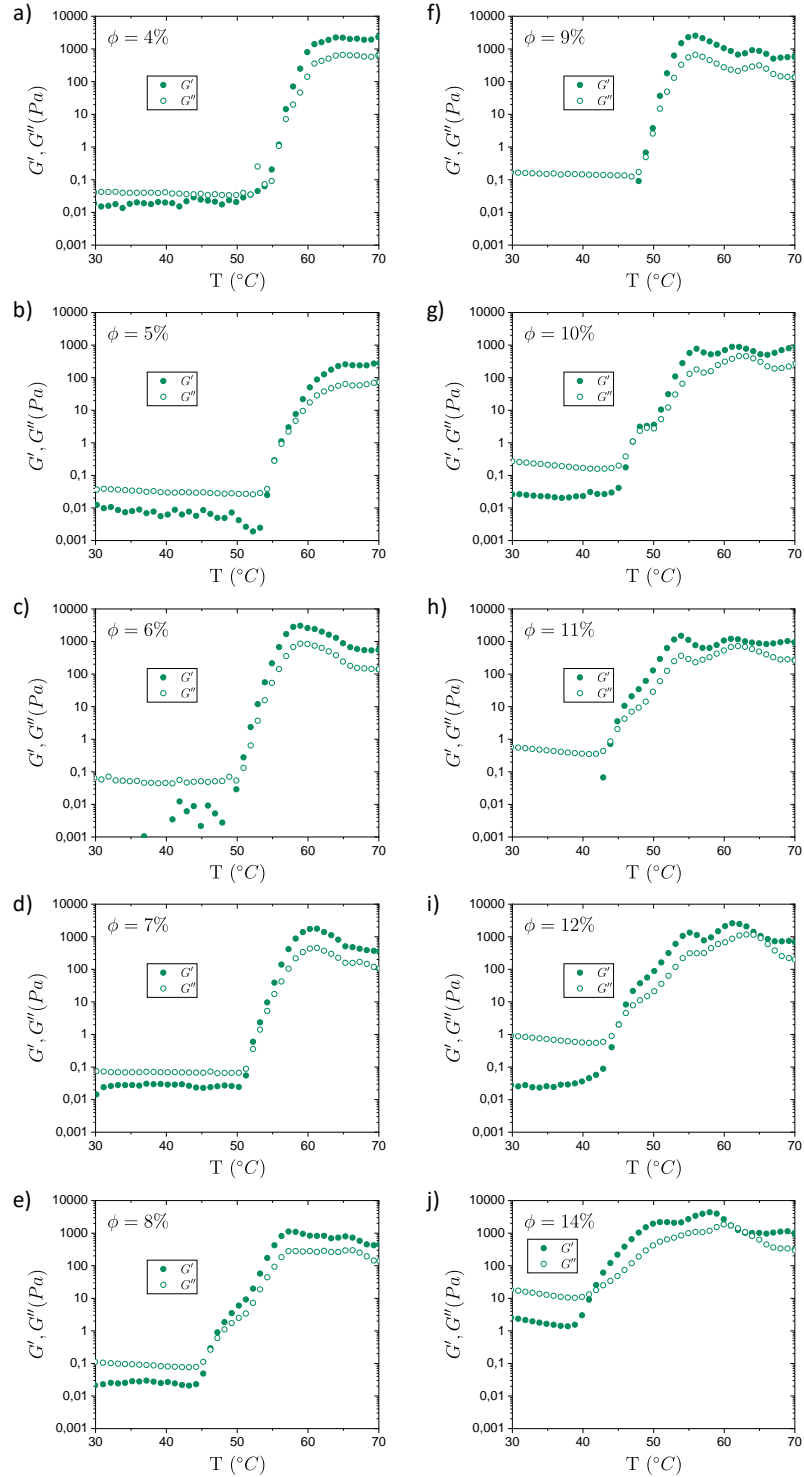

**Figure S7**. Evolution of the elastic (G', full symbol) and loss (G'', empty symbol) moduli as a function of temperature for coated nanoplatelet suspensions at different mass concentration ranging from 4 wt% to 14 wt% as indicated in the inset: a) $\Phi = 4\%$ ; b) $\Phi = 5\%$ ; c) $\Phi = 6\%$ ; d) $\Phi = 7\%$ ; e) $\Phi = 8\%$ ; f) $\Phi = 9\%$ ; g) $\Phi = 10\%$ ; h) $\Phi = 11\%$ ; i) $\Phi = 12\%$ ; j) $\Phi = 14\%$.

## 13. Macroscopic state of the laponite suspensions after several years.

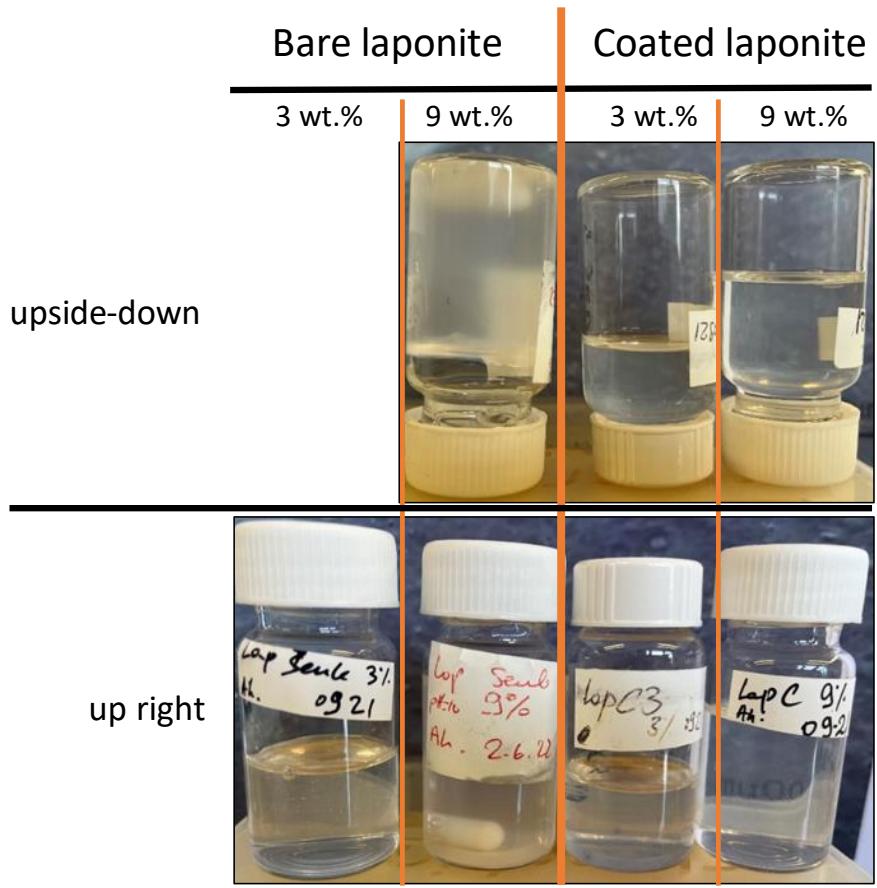

**Figure S8**. Photos taken at room temperature of bare laponite and coated laponite suspensions after being stored for several years at 4°C. The most recent sample dates back to June 2022 (i.e. bare laponite at 9 wt.%), while the oldest dates back to September 2021 (i.e. bare laponite at 3 wt.% and the two samples of coated laponite). The 3 samples in the upper image are turned upside-down

## 14. Estimation of ionic strength and Debye length.

The samples were prepared by mixing the solid laponite powder with aqueous solutions at fixed pH = 10 with NaOH. The nanoplatelets bear both negative charges on the faces and positive charges on the rims. The number of negative charges per nanoplatelet have been measured for different concentrations and found to be about 750 e. The number of the rim charges is more difficult to measure, however the typical estimate is roughly 10% of the negative charges [B. Ruzicka, and E. Zaccarelli, A fresh look at the Laponite phase diagram. Soft Matter, 2011, 7, 1268]. We used the value of 75 rim charges per nanoplatelet to determine the concentration of excess (4Na$^+$, P$_2$O$_7^{4-}$) salt in each solution knowing that 75 pyrophosphate anions are attached to each nanoplatelet and the remaining acts as added salt that constitutes the double layer. We considered a total pyrophosphate concentration of 8.2 g per 100 g of dry laponite powder, as indicated by the supplier. The ionic strength, $I = \frac{1}{2}\sum_i c_i Z_i^2$, and thus the Debye length, $\kappa^{-1}$, of the suspensions could be estimated by taking into account the contribution of Na$_4$P$_2$O$_7$ (i.e. 4Na$^+$, P$_2$O$_7^{4-}$) in excess (i.e. that don't interact with the positive charges on the rims). Thus, we assume that Na$^+$ counter ions of the laponite platelets do not contribute significantly to the ionic strength. [A. Mourchid et al., Langmuir 1995, 11, 1942 ; M. Dubois et al. J. Chem. Phys. 1992, 96, 2278]. In our experiments, we estimate that the variation in laponite mass fraction ($\phi$) from 1% to 14% is associated with a variation in Debye length (i.e. $\kappa^{-1} = \sqrt{\left(\frac{\varepsilon_r \varepsilon_0 kT}{2e^2} \times \frac{1}{I}\right)}$ with I the ionic strength calculated by only taking into account the Na$_4$P$_2$O$_7$ in excess) from approximately 2.3 to 0.6 nm (Table S9).

The concentration of each species (i.e. nanoplatelet, copolymer, total pyrophosphate present in the raw material and calculated free pyrophosphate in each solution) are given in the table S9 with the calculated ionic strength, I, and Debye length, $\kappa^{-1}$:

| Nanoplatelets (wt. %) | Polymer (wt. %) | Anions (M) | Rim charges (M) | Free anions* (M) | Ionic strength** (M) | Debye length*** (Å) |
|---|---|---|---|---|---|---|
| 1 | 0.81 | 0.0031 | 0.0010 | 0.0021 | 0.017 | 22.7 |
| 2 | 1.62 | 0.0062 | 0.0021 | 0.0041 | 0.035 | 16.0 |
| 3 | 2.43 | 0.0093 | 0.0031 | 0.0062 | 0.052 | 13.1 |
| 4 | 3.24 | 0.0124 | 0.0042 | 0.0082 | 0.070 | 11.3 |
| 5 | 4.05 | 0.0155 | 0.0052 | 0.0103 | 0.087 | 10.1 |
| 6 | 4.86 | 0.0186 | 0.0062 | 0.0123 | 0.105 | 9.3 |
| 7 | 5.67 | 0.0217 | 0.0073 | 0.0144 | 0.122 | 8.6 |
| 8 | 6.48 | 0.0248 | 0.0083 | 0.0165 | 0.140 | 8.0 |
| 9 | 7.29 | 0.0279 | 0.0094 | 0.0185 | 0.157 | 7.6 |
| 10 | 8.1 | 0.0310 | 0.0104 | 0.0206 | 0.175 | 7.2 |
| 11 | 8.91 | 0.0341 | 0.0114 | 0.0226 | 0.192 | 6.8 |
| 12 | 9.72 | 0.0372 | 0.0125 | 0.0247 | 0.210 | 6.6 |
| 13 | 10.53 | 0.0403 | 0.0135 | 0.0268 | 0.227 | 6.3 |
| 14 | 11.34 | 0.0434 | 0.0146 | 0.0288 | 0.245 | 6.1 |

**Table S9.** Table summarizing the evolution of the concentration of the various components of the system and of the Debye length as a function of the concentration of laponite.

\* We assume that the free anions are only related to the contribution of the Na$_4$P$_2$O$_7$ (i.e. 4 Na$^+$ + P$_2$O$_7^{4-}$) in excess.
\*\* I = 0.5\*([ Na$^+$]\*1+[ P$_2$O$_7^{4-}$]\*4$^2$) = 8.5 \* [Na$_4$P$_2$O$_7$]$_{excess}$= 8.5 \* [Free anions].

*** $\kappa^{-1} = \frac{0.304}{\sqrt{I}}$ in water at 20°C.

We also measured at 20°C the evolution of the conductivity of the coated Laponite suspension as a function of the Laponite concentration as shown in figure S10. The measurement were done with a NanoZS apparatus (Malvern Instrument) using a standard folded capillary zeta cell. The observed variation of conductivity qualitatively reflects the expected effect of the ionic strength.

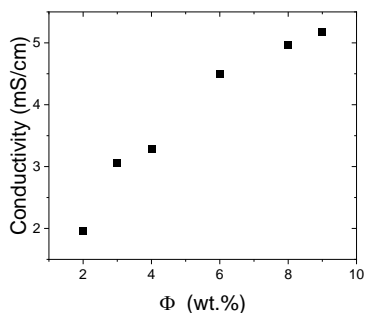

**Figure S10.** Evolution of the conductivity of the coated laponite suspensions as a function of the laponite concentration (T = 20°C).